\documentstyle[referee,epsf]{mn}

\newcommand\eq{\begin{equation}}
\newcommand\en{\end{equation}}

\begin{document}

\title{A Comparison of Simulated and Analytic Major Merger Counts}
\author[Cohn, Bagla \& White]{
J. D. Cohn${}^1$, J. S. Bagla${}^{1,2}$, Martin White${}^1$\\
${}^1$Harvard-Smithsonian Center for Astrophysics, 60 Garden St., 
Cambridge, MA 02138, USA\\
${}^2$Mehta Research Institute, Chhatnag Road, Jhusi, Allahabad
211019, India}
\date{September 2000}

\maketitle

\begin{abstract}
We use large volume, high resolution, N-body simulations of  3 different
$\Lambda$CDM models, with different clustering strengths, to generate dark
matter halo merging histories.  Over the reliable range of halo masses,
roughly galaxy groups to rich clusters of galaxies, we quantify the number
density of major mergers for two different time intervals and compare with
analytic predictions based on the extended Press-Schechter~\shortcite{ps}
theory.
\end{abstract}


\section{Introduction}

It is now widely accepted that the observed large-scale structure in the
universe formed hierarchically, through the process of gravitational
amplification of small initial perturbations in a universe with
predominantly cold dark matter (CDM).
Within this ``bottom-up'' paradigm, larger structures form by the merging
of smaller units, resulting in a dynamic and continually evolving
matter distribution.  
Multiple lines of evidence support this scenario,
suggesting that clusters and groups are still
forming in the present universe, and that the idealization of a relaxed,
virialized structure is somewhat unrealistic.
Thus the merger history is of fundamental importance in determining not
simply the details of structure formation but also in many cases 
the global properties of clustered objects
under consideration.

Among merger events there are two extreme limits: major mergers and accretion.
In a major merger, the masses of the progenitor halos are comparable, and
consequently their interaction results in a halo which is dynamically
disrupted for some time after the merger.
In contrast accretion is the merger of a small halo with a much larger
halo, which is generally not severely disrupted by the process.
We shall focus throughout on major mergers of objects ``in the field''
(which are expected to differ from major mergers of sub-haloes
within a larger halo, e.g. ~galaxies in a galaxy cluster,
Cavaliere \& Menci~\shortcite{cava1}).  

The dynamical disruption caused by a major merger is expected to have many
observational consequences
(for an extensive list see Roettiger et al.~\shortcite{roettiger}).
For mergers of groups and clusters, these include among other things:
inducing scatter in IGM temperature \cite{mathie},
increasing and then decreasing star formation rates (e.g.
~Fujita et al~\shortcite{fujita}, for galaxy size halo consequences
see ~Barnes and Hernquist~\shortcite{barnes},
and references therein),
disrupting cooling flows (e.g.~Allen et al.~\shortcite{allen}) and
producing non-thermal radiation \cite{blasi}.

For all of these reasons there has been intense study of mergers analytically,
numerically and observationally.  On the theoretical side these range from
in depth studies of specific mergers \cite{huss}
to properties of mergers as an ensemble, their rates, consequences for
formation times, etc.,
e.g.~in Bond et al.~\shortcite{bondetal}, Lacey \& Cole \shortcite{lc93,lc94},
Kitayama \& Suto \shortcite{kit-sut,kit-sut2}, Tormen \shortcite{tormen},
Somerville et al.~\shortcite{sometal},
Percival \& Miller \shortcite{permil}.
This paper is an instance of the latter, that is, a study of the bulk
statistical properties of major mergers in the field.
Specifically the question we address is:
What is the number density of objects, at a given time and mass, that
have undergone a major merger within the last 0.5 or 2.5 Gyr?
Part of our aim is to address this question with several different methods,
to further understand our answers and to investigate the accuracy of some
analytic estimates based on ``extended'' Press-Schechter~(\shortcite{ps};
hereafter PS) theory (see below).

The outline of the paper is as follows.  
We discuss the simulations in \S\ref{sec:sims} and the numerical merger
counts in \S\ref{sec:counts}.  A review of some of the relevant analytic
work is given in \S\ref{sec:analytic} and these estimates are compared with
our numerical results in \S\ref{sec:compare}.  Some technical details
are relegated to an Appendix.

\section{Simulations} \label{sec:sims}

We have chosen to focus on high mass objects,
roughly galaxy groups to cluster scales, where we believe that the
evolution is dominated by gravity and thus relatively inexpensive to
compute numerically and under reasonable theoretical control.
Our choice of time-scales is motivated by the fact that 0.5 Gyr is close
to that relevant for star formation caused by mergers (e.g.
as noted in Fujita et al~\shortcite{fujita} and
for galaxies argued by Percival \& Miller~\shortcite{percetal}
using luminosity curves in Bruzual \& Charlot\shortcite{starform}) while
the larger time interval, 2.5 Gyr, is closer to those expected for cluster
relaxation processes (e.g.~see Mathiesen \& Evrard~\shortcite{mathie}).

We ran three simulations of a $\Lambda$CDM model with a high resolution
N-body code \cite{treepm}.
The cosmological models all assumed $\Omega_m=0.3$ and
$\Omega_\Lambda=0.7$ but differed in $z=0$ clustering strengths.
We generated one realization for each of $\sigma_8=0.8$, 1.0 and 1.2,
where $\sigma_8^2$ is the variance of the matter fluctuations in top-hat
spheres of radius $8 h^{-1}$Mpc.
Each simulation employed $128^3$ dark matter particles in a box of side
$256\,h^{-1}$Mpc, i.e.~a particle mass of $6.7\times 10^{11}h^{-1}M_\odot$,
with a Plummer force law with softening length $200h^{-1}$kpc.
The initial conditions were generated by displacing particles from a regular
grid, using the Zel'dovich approximation,
with the initial 
redshift was chosen so that the
maximum displacement was (slightly) smaller than the mean interparticle spacing
due to the large volume of the simulation.
Time steps were chosen to be a small fraction of the shortest dynamical time of
any particle in the simulation.
The initial redshift and number of steps ranged from
$z_{\rm init}=9$ and 445 time steps for $\sigma_8=0.8$,
to $z_{\rm init}=14$ and 635 time steps for $\sigma_8=1.2$.
Further details (transfer function used, etc.) are given in the Appendix.

The large volume of the simulations provides good statistics
on the high-mass end of the mass function, where we have focused our
attention.
We have checked that the resulting mass functions scale as expected over
the range of masses and redshifts which shall be of interest in this work.
In our merger counts we do notice some possible effects at early times for
low mass halos in the $\sigma_8=0.8$ simulation, and these would be
consistent with residuals from our grid initial conditions and limited
numerical resolution.  We shall comment further on this in \S\ref{sec:compare}.

\subsection{Group catalogues}

The full particle distribution was dumped every 0.5 Gyr starting from $z=2.16$.
For each output we generated a group catalogue using one of two group
finders.  We used HOP \cite{hop} for all three runs.
For comparison, as well as to make explicit contact with other work, we
additionally used the friends-of-friends algorithm
(FOF\footnote{The implementation used here is from
http://www-hpcc.astro.washington.edu/tools/FOF/ .},
Davis et al.~\shortcite{fof})
for the $\sigma_8=1$ run.
{}From these catalogues we were able to construct merger trees (see below)
back to $z=1.86$ for 0.5 Gyr intervals and back to redshift $z=1.11$ for
2.5 Gyr intervals.

Since it may be unfamiliar, we briefly review the operation of HOP.  For
technical details the reader is referred to Eisenstein \& Hut~\shortcite{hop}.
HOP finds groups by first assigning each particle a density, and then
``hopping'' to neighbors with a higher density.  Each particle belongs to the
same class as its densest neighbor, and in this way each particle is assigned
to a local density peak.
To correct for the possibility of local density maxima causing groups to
fragment, groups are merged if the bridge between them exceeds some chosen
density threshold.
There are 6 parameters one must choose for HOP: we used the default values
for $N_{\rm merge}$ and $N_{\rm hop}$ (number of neighbors to look at when
searching for a boundary and the densest neighbor) but took a lower value
of the number of particles averaged over when calculating the density
($n_d=8$ rather than $64$).  The results were similar for $n_d=8$ and
$n_d=16$, with the higher $n_d$ excluding small groups.
We excluded all groups of fewer than 8 particles.
The remaining three parameters are all density thresholds.
Eisenstein \& Hut~\shortcite{hop} claim the method works best if the thresholds
are in the ratio $1:2.5:3$, so we took
$\delta_{\rm outer}=50$, $\delta_{\rm saddle}=125$ and $\delta_{\rm peak}=150$.
These thresholds correspond to: the required density for a particle to be in
a group, 
the minimum boundary density between two groups for them to be merged,
and the minimum central density for a group to be independently viable.
We experimented with these thresholds and found negligible differences in the
final catalogues.
At redshift zero there were between $18,000$ and $19,000$ groups
(see Fig.~\ref{fig:hz0}).

The Friends-of-Friends algorithm is much better known.  It has only a single
free parameter: the linking length $b$, usually given in units of the mean
interparticle spacing.  FOF defines groups of particles which are each
separated by less than $b$ from at least one other member of the group.
Roughly speaking a FOF group consists of all particles within an iso-density
contour of $b^{-3}$ of the background matter density.
We follow convention and choose  $b=0.2$ (corresponding to a local
overdensity $125$). 
For the $\sigma_8=1$ case studied here, some notable differences for
merger counts and number counts were found, shown in Fig.~\ref{fig:hopfof}.
For halos with more than 50 FOF particles, FOF halos had on average
about 90 per cent of the particles of the corresponding HOP halos.
FOF did find a larger total number halos, 
but these extra halos were at very low mass. 

Both of these group finders do not identify substructure within large
halos.  This is not a particular problem for our purposes since we are
interested primarily in mergers which take place in the field.
Mergers within clustered environments are expected to behave differently
\cite{cava1} and have been studied by other groups, e.g. 
Gottloeber, Klypin \& Kravtsov \shortcite{gottl}.

\subsection{Merger trees}

The definition of a major merger is somewhat arbitrary as there is
some dependence on what physical consequence is of interest. 
In this paper we define major mergers to be 
mergers in which the portions of the
two largest halos that become part of the final halo
have a mass ratio between 1:1 and 1:5.\footnote{Note that this definition
also allows an unbound collection of particles from a parent halo 
to count as a predecessor for
the purposes of defining a merger.  Quantifying how closely this criterion
restricts one to a resulting dynamically disrupted halo would be
very interesting.  We thank the referee for emphasizing this issue.} 
We compute the number of mergers within a given time interval,
a very interesting quantity from an observational and theoretical standpoint.
Note that this is an inclusive rate  -- the number of mergers is not given
directly by a merger rate since in a given time interval, halos can merge and
then accrete, or accrete, merge, accrete, or merely merge, etc.
We count any halo which has a major merger in the given time interval,
regardless of what else occurred to it during the interval.

Having obtained the group catalogues for each output we constructed a merger
tree for each simulation.
For each time and for each (parent) group, we identified the (daughter) group
membership of all the constituent particles at the next output time.
The parent group was then considered a predecessor of the daughter group
if more than 8 parent-group particles belonged to the daughter group at the
next output time.
For 0.5 Gyr merger trees, four of these ``daughter groups'' were considered
for each predecessor group.  This resulted in missing very few of the
daughter groups~--~less than twenty total daughter groups
went uncounted for the combined 22 outputs
times three values of $\sigma_8$ for HOP (there were $\gg$ 10,000 groups 
for most time steps).
For the $2.5$ Gyr merger trees, a more conservative maximum of 15 daughters
was recorded for each predecessor group and no predecessor group had more
than this number.

\section{Numerical merger counts} \label{sec:counts}

The runs and merger tree construction described above resulted in 
halo evolution and merger counts with the following properties.
We found, in agreement with Tormen~\shortcite{tormen}, that about 20 per
cent of the mass of the halos was lost as the halos evolve.  In fact some
groups, in particular many of those close to the minimum group size,
disappeared completely between time steps.
While this could be partly due to our finite force resolution, it also
highlights the ambiguity inherent in halo identifications using these
group finders. 

We anticipate that small mass predecessors may be undercounted, especially at
earlier times.
For this reason we only present results for halos with more
than $10$ times the minimum group size, or $5\times 10^{13}h^{-1}M_\odot$.
\footnote{This was because one could find in
the $0.5$ Gyr merger trees major mergers with a second predecessor as
small as $1/10$ the size of the final mass.  For the $2.5$ Gyr merger
trees of order $20$ of the second predecessors were smaller than this 
fraction for medium range masses, for each value of $\sigma_8$, summed
over all times.}
When predecessor sizes are compared for the purpose
of identifying a major merger (i.e.~a ratio $1:1$ to $1:5$), the
number of particles from the predecessor group that go into the final
group, not the number of total particles in the predecessor group are
considered.  

\begin{figure}
\begin{center}
\leavevmode
\epsfxsize=3.2in\epsfbox{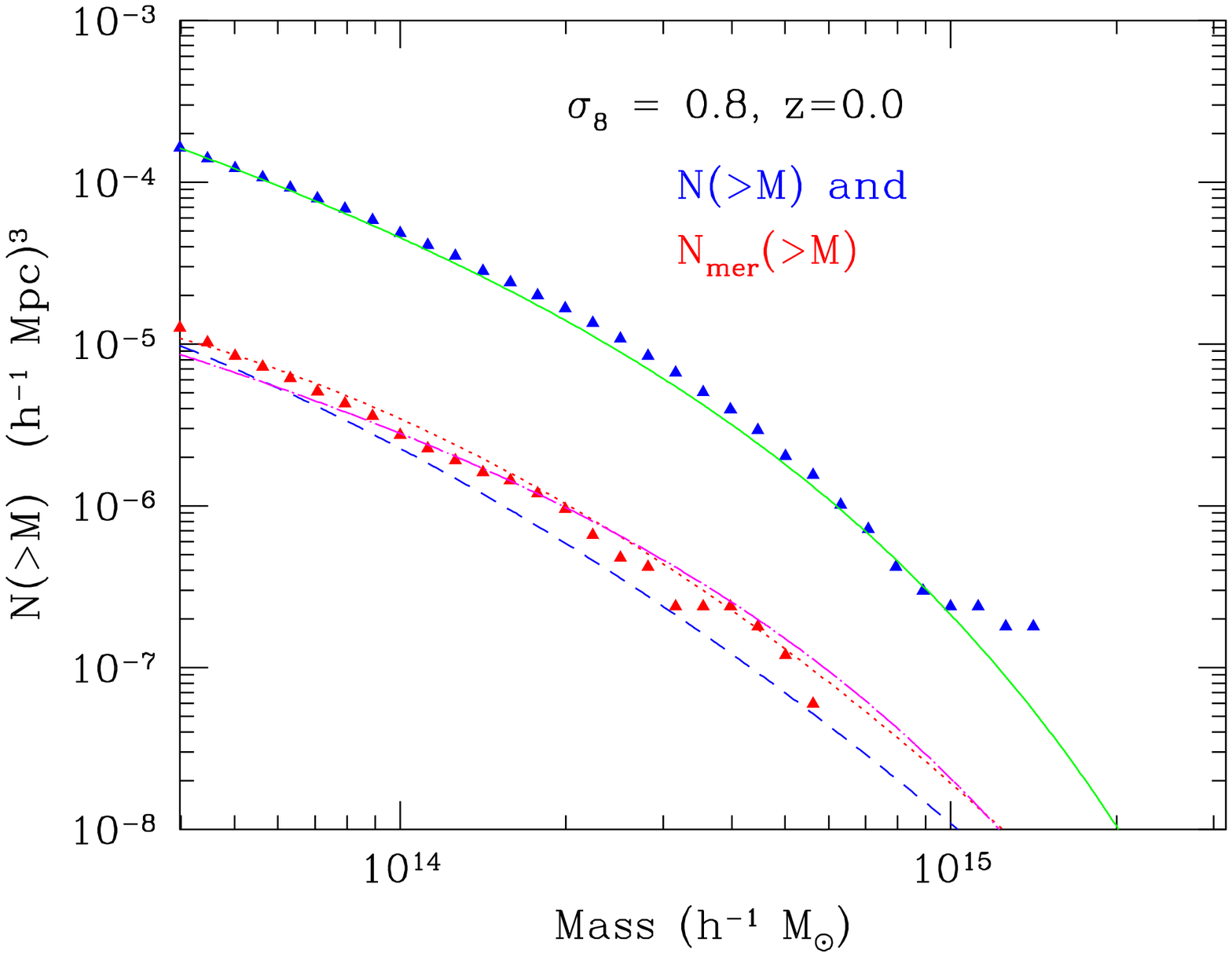}
\epsfxsize=3.2in\epsfbox{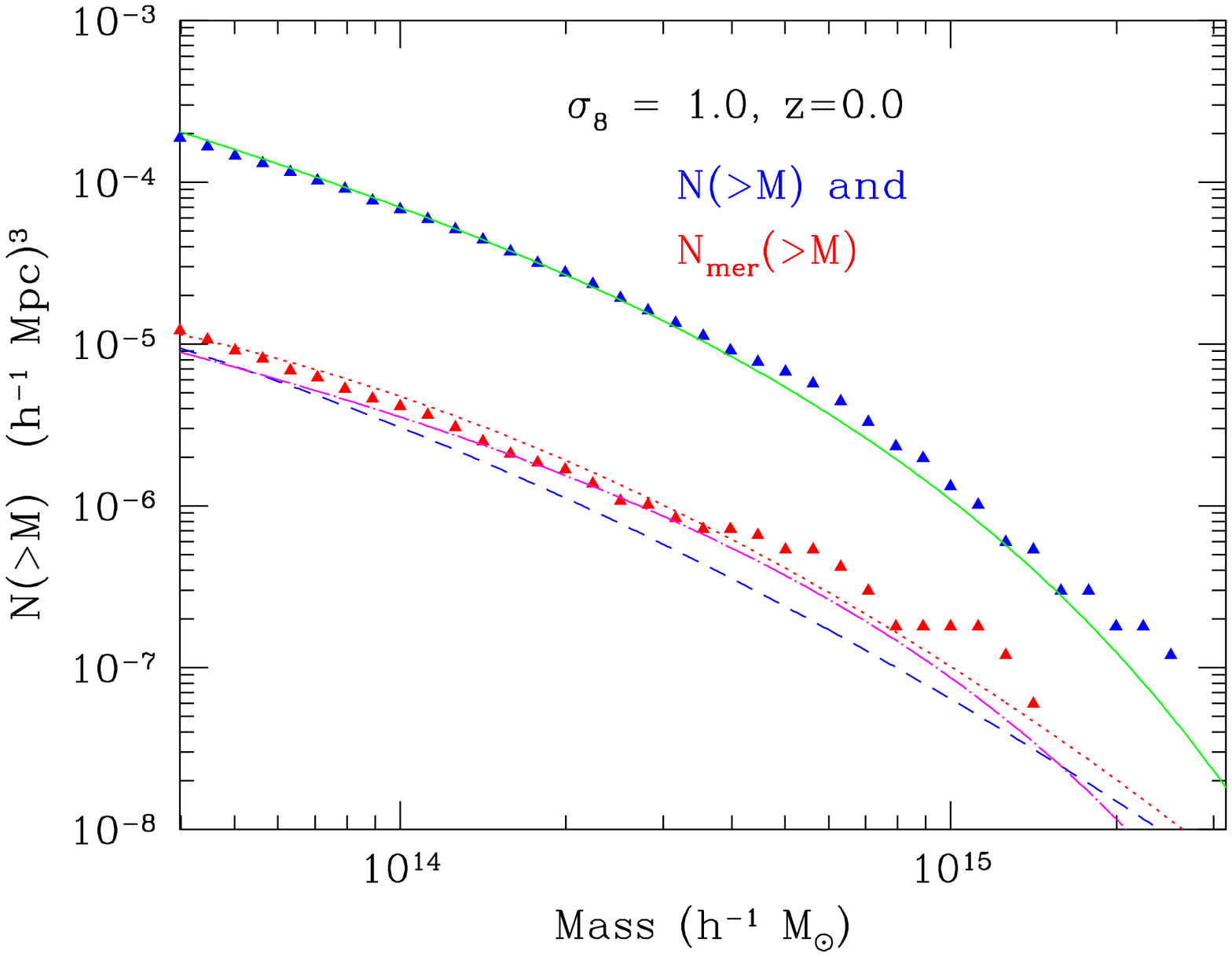}
\\
\epsfxsize=3.2in\epsfbox{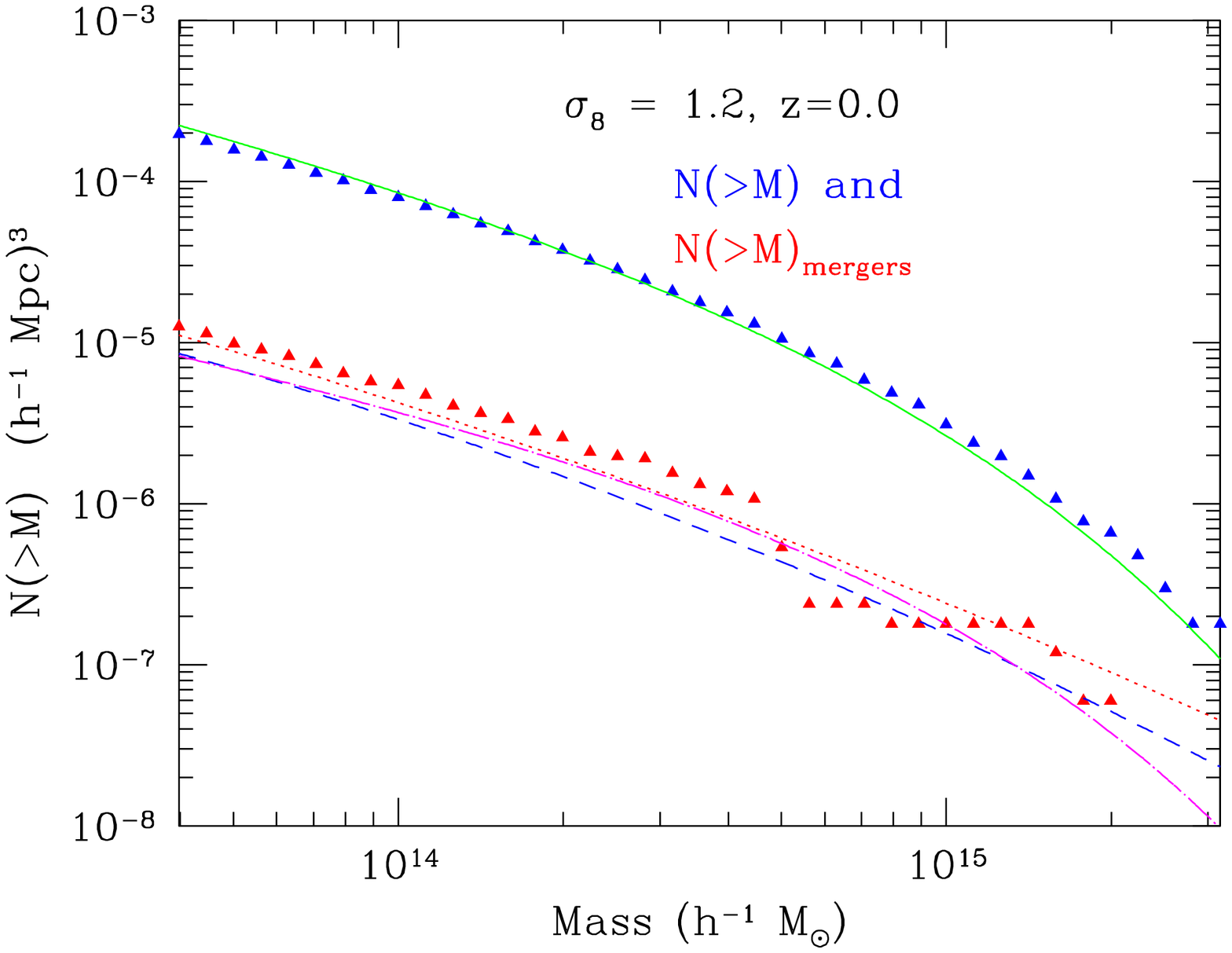}
\end{center}
\caption{Number density of groups and those with recent (within
$0.5$ Gyr) major mergers at $z=0$ found by HOP, for three values of 
$\sigma_8$.  The triangles are the simulations, the upper (smooth) curve
is the PS prediction (Eq.~\protect\ref{nps}) with $\delta_c=1.48$.
The lower curves are quadratic fits as a
function of $\log $ mass and $\ln (1+z)$.  The dotted line is the
fit to the simulation results, the dashed line
corresponds to the fit to the ``direct'' calculation,
Eq.~(\protect\ref{mercalc}),
and the dot-dashed line corresponds to the ``jump'' estimate given in
Eqs.~(\protect\ref{psmer}).  One halo in the entire
simulation volume corresponds to a density of $6 \times 10^{-8}
(h^{-1} {\rm Mpc})^3$.}
\label{fig:hz0}
\end{figure}

The number density of halos and of merger counts (within $0.5$ Gyrs)
at redshift zero for the three values of $\sigma_8$ (using HOP), are
given in Fig.~\ref{fig:hz0}.
The smooth curve going through the upper set of points (the number density)
is the PS fit discussed in the next section. 
As we can see, the PS theory provides an adequate fit to the simulation
results, as has been noted before by numerous authors
\cite{efstat,efrees,wef,lc94,ecf,gross,governato}.
The lower curves correspond to two of the analytic fits, also discussed in
the next section, as well as a quadratic fit to $\log M/(h^{-1}M_\odot)$ and
$\ln (1+z)$ for all times.

\begin{figure}
\begin{center}
\leavevmode
\epsfxsize=3.2in\epsfbox{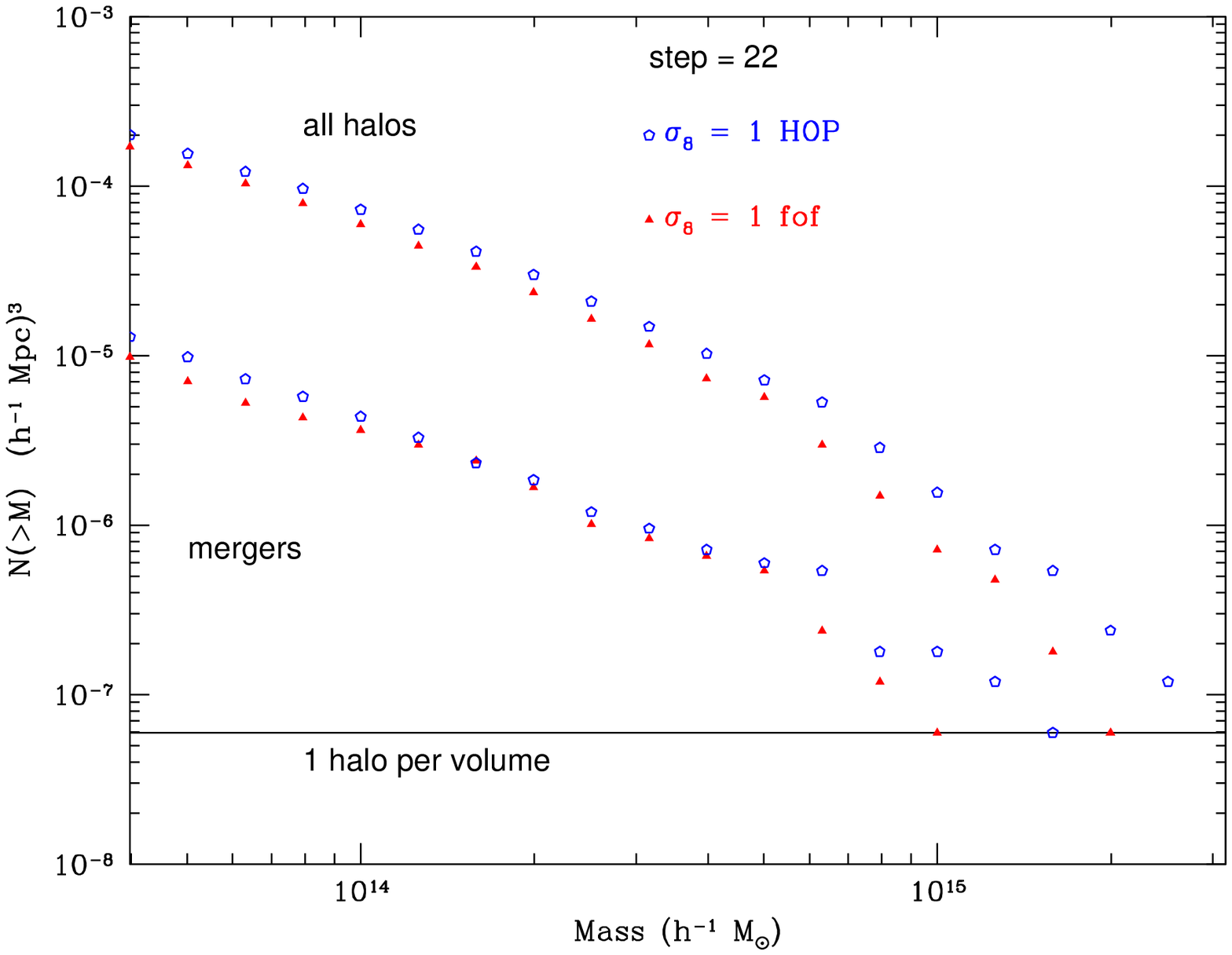}
\epsfxsize=3.2in\epsfbox{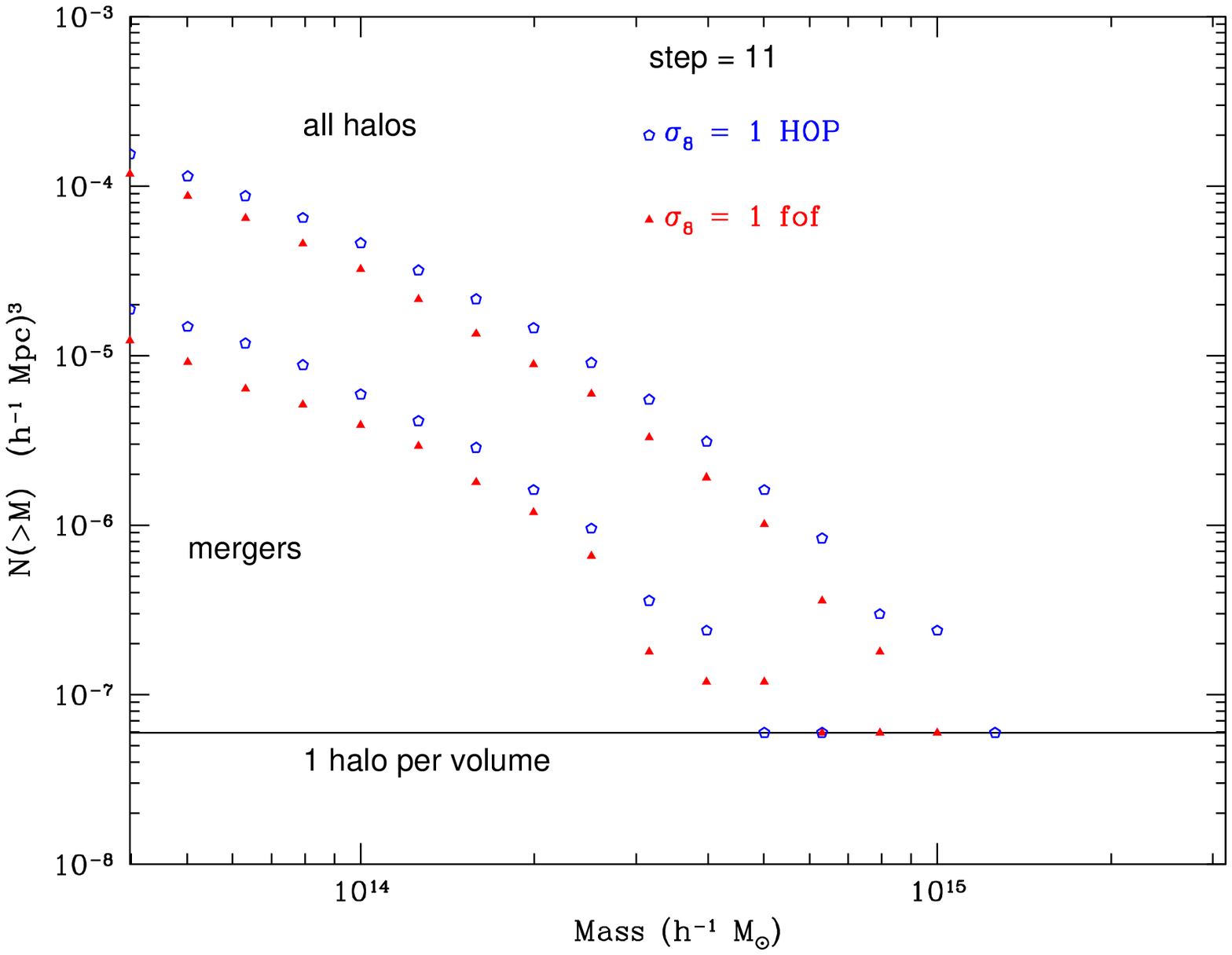}
\end{center}
\caption{Number densities of halos and mergers for the
0.5 Gyr time interval, using two group
finders, HOP and FOF, on the $\sigma_8=1$ simulation.
At left is redshift $z=0$, at right is $z = 0.53$.  The best fitting
$\delta_c$ in the PS formalism for the HOP halo number
density is $\delta_c = 1.48$ and for the FOF halo
number density is $\delta_c = 1.64$.}
\label{fig:hopfof} 
\end{figure}

Many groups have found dependence on group finder of their results,
e.g. comparing SO(178) and FOF with different linking lengths
(e.g. Lacey and Cole \shortcite{lc94}, Tormen \shortcite{tormen}). 
Our FOF merger trees differed from those for halos identified with HOP, 
just as the total
numbers of halos differed.
The comparison between the HOP and FOF groups and merger trees can
be seen in Fig.~\ref{fig:hopfof}.

\begin{figure}
\begin{center}
\leavevmode
\epsfxsize=3.2in\epsfbox{s1hz0.ps}
\epsfxsize=3.2in\epsfbox{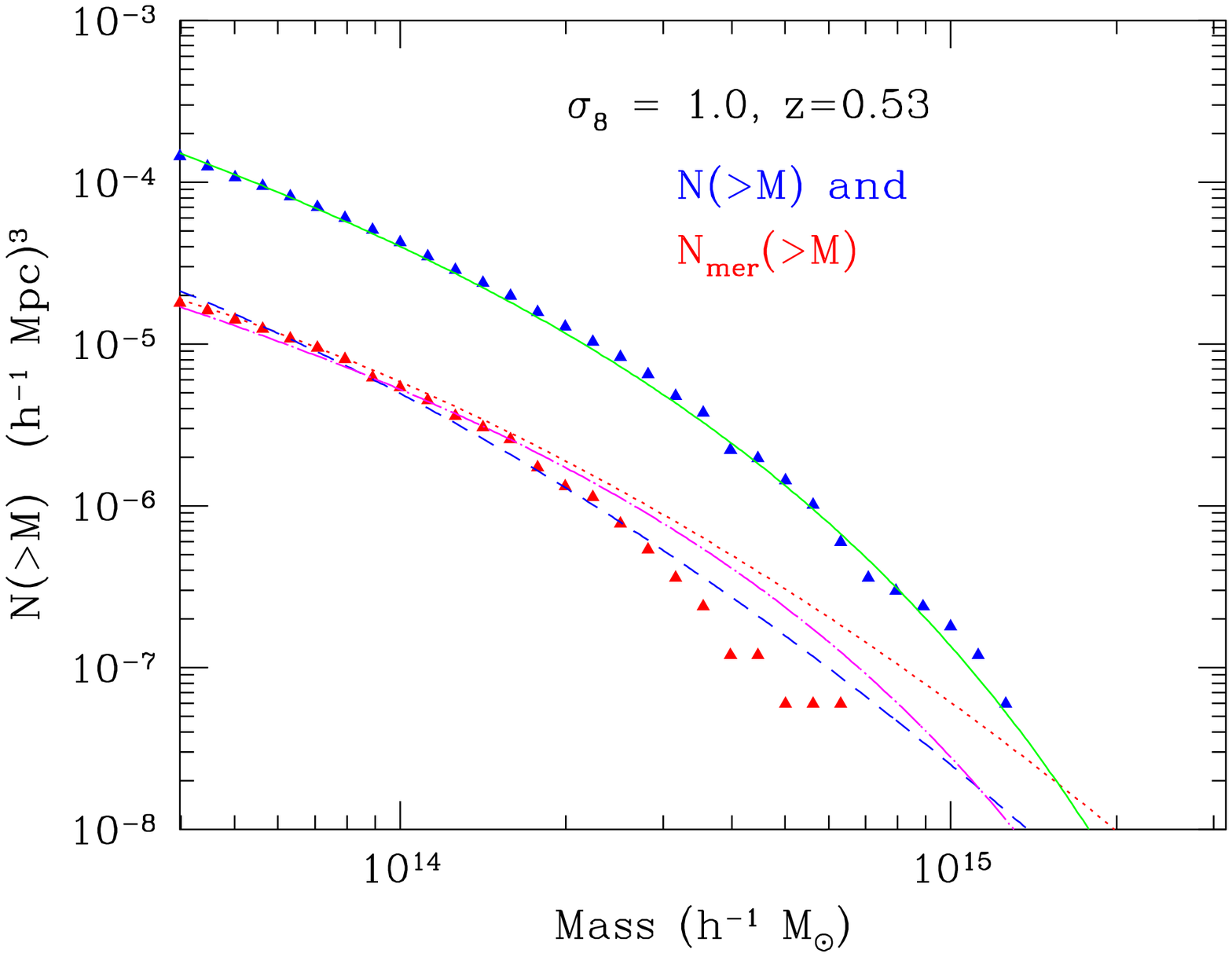}
\\
\leavevmode
\epsfxsize=3.2in\epsfbox{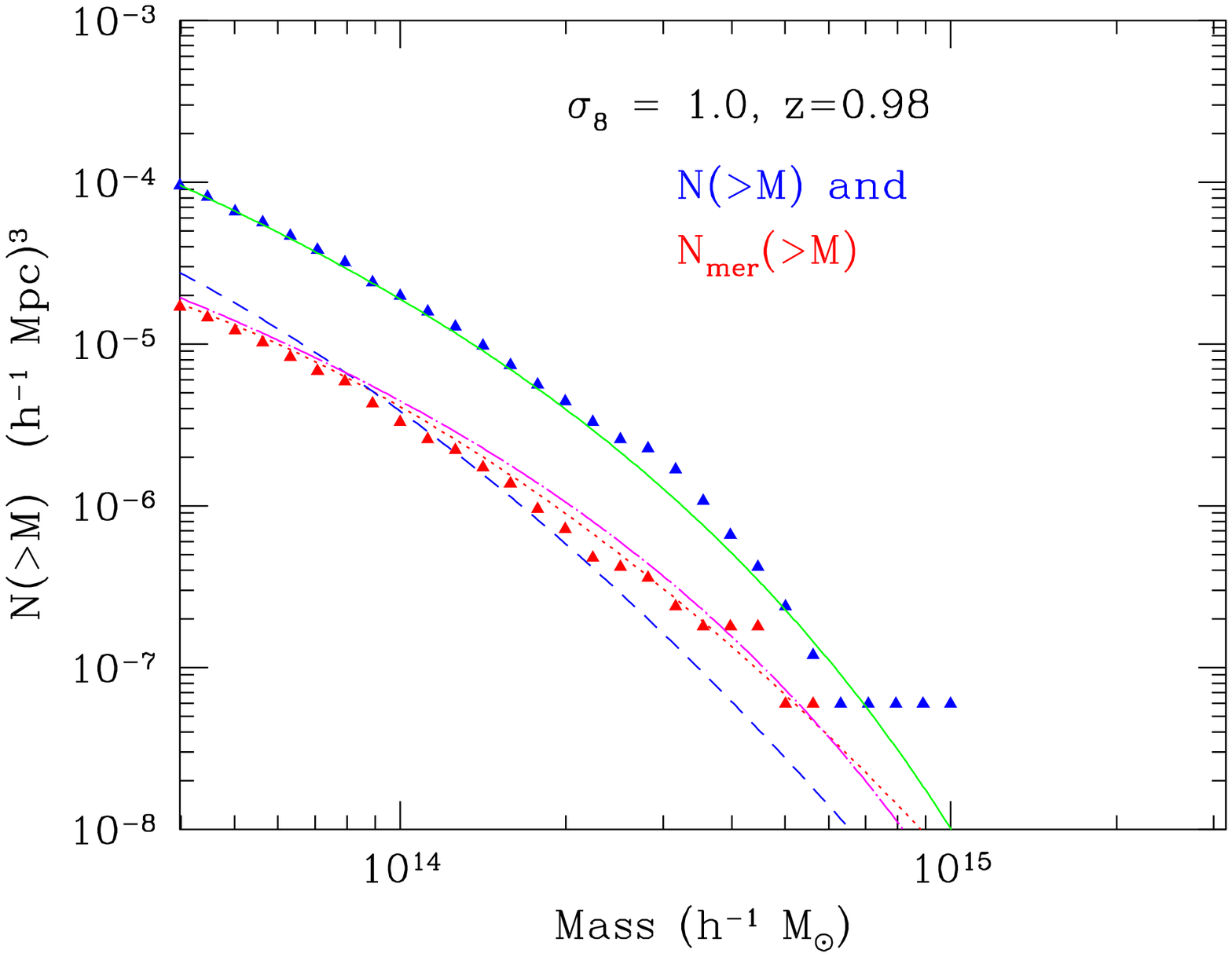}
\epsfxsize=3.2in\epsfbox{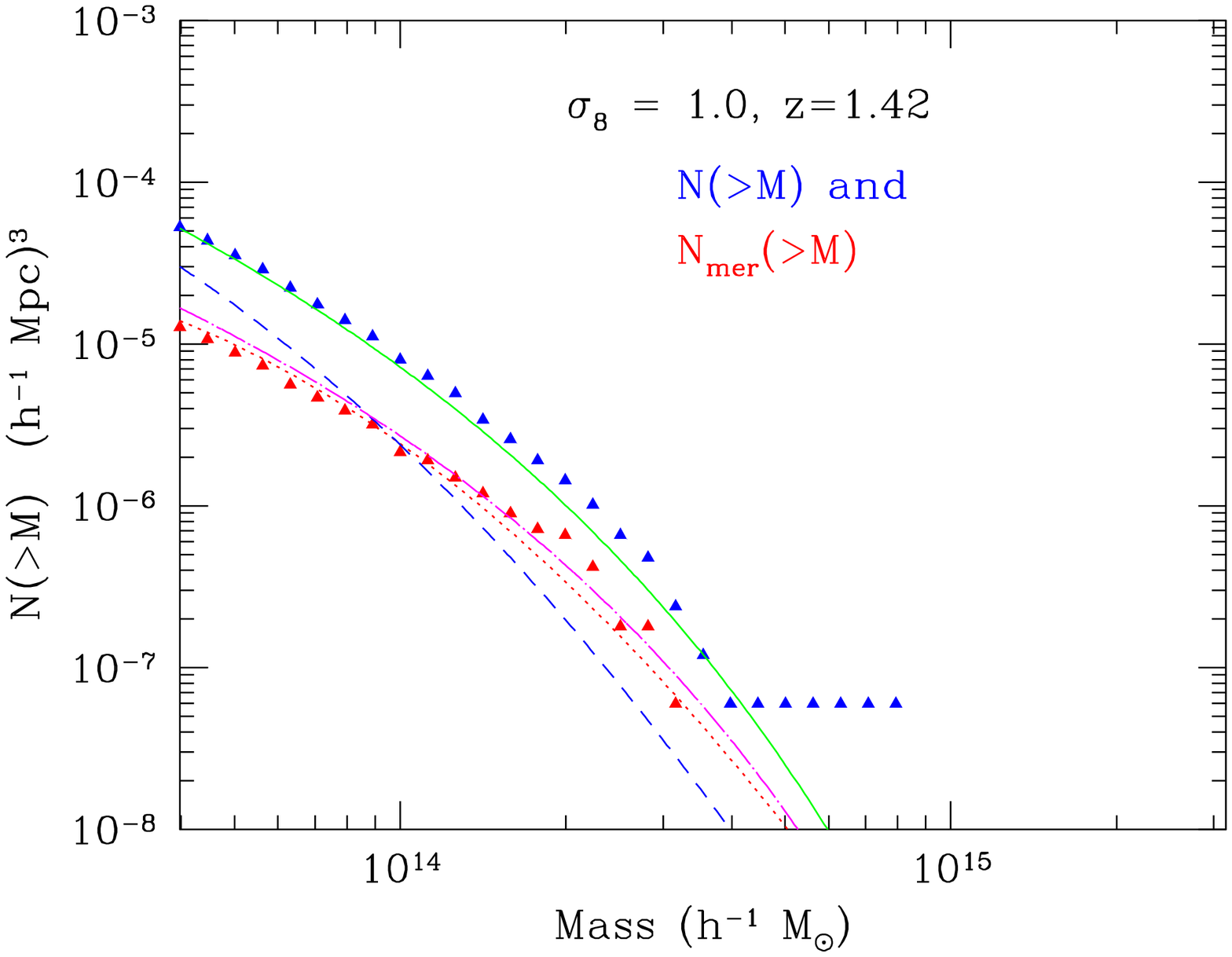}
\end{center}
\caption{Number density of groups and major mergers within $0.5$ Gyrs
at $z=0.0, 0.53, 0.98, 1.42$ for $\sigma_8 = 1$.  The curves are as in
Fig.~\protect\ref{fig:hz0}, i.e., the upper (smooth) curves again are PS
predictions for number densities with $\delta_c = 1.48$ and the lower
curves are quadratic fitting functions (as a function of $\log $ mass
and $\ln (1+z)$) to merger counts in section \S3: the dotted line is
the fit to the simulation results, the dashed line corresponds to the fit to
the ``direct'' calculation, Eq.~(\protect\ref{mercalc}), and the
dot-dashed line corresponds to the ``jump'' estimates given in
Eqs.~(\protect\ref{ksform},\protect\ref{psmer}).  One halo in the entire
simulation volume corresponds to a density of $6 \times 10^{-8}
(h^{-1} {\rm Mpc})^3$.} 
\label{fig:hz1}
\end{figure}

One can consider the number of mergers within a given recent time
interval, taken here to be 0.5 Gyr,
as a function of redshift, as is shown in Fig.~\ref{fig:hz1}.
One can see that the number of high mass halos decreases as one goes back
to earlier times, and that simultaneously the fraction of total halos of
any given mass which have undergone a major merger within a recent
fixed time interval increases.

\begin{figure}
\begin{center}
\leavevmode
\epsfxsize=3.2in\epsfbox{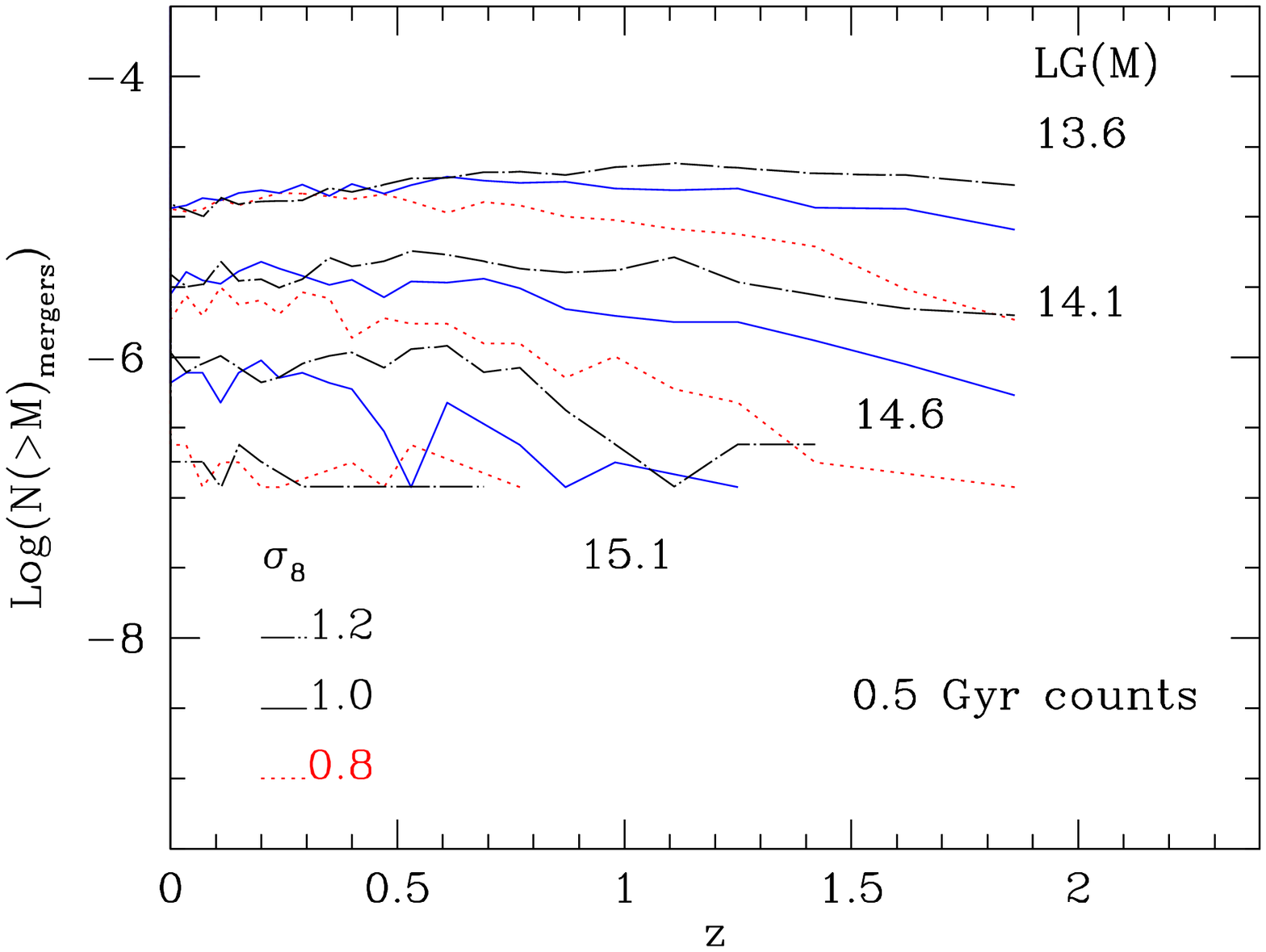}
\epsfxsize=3.2in\epsfbox{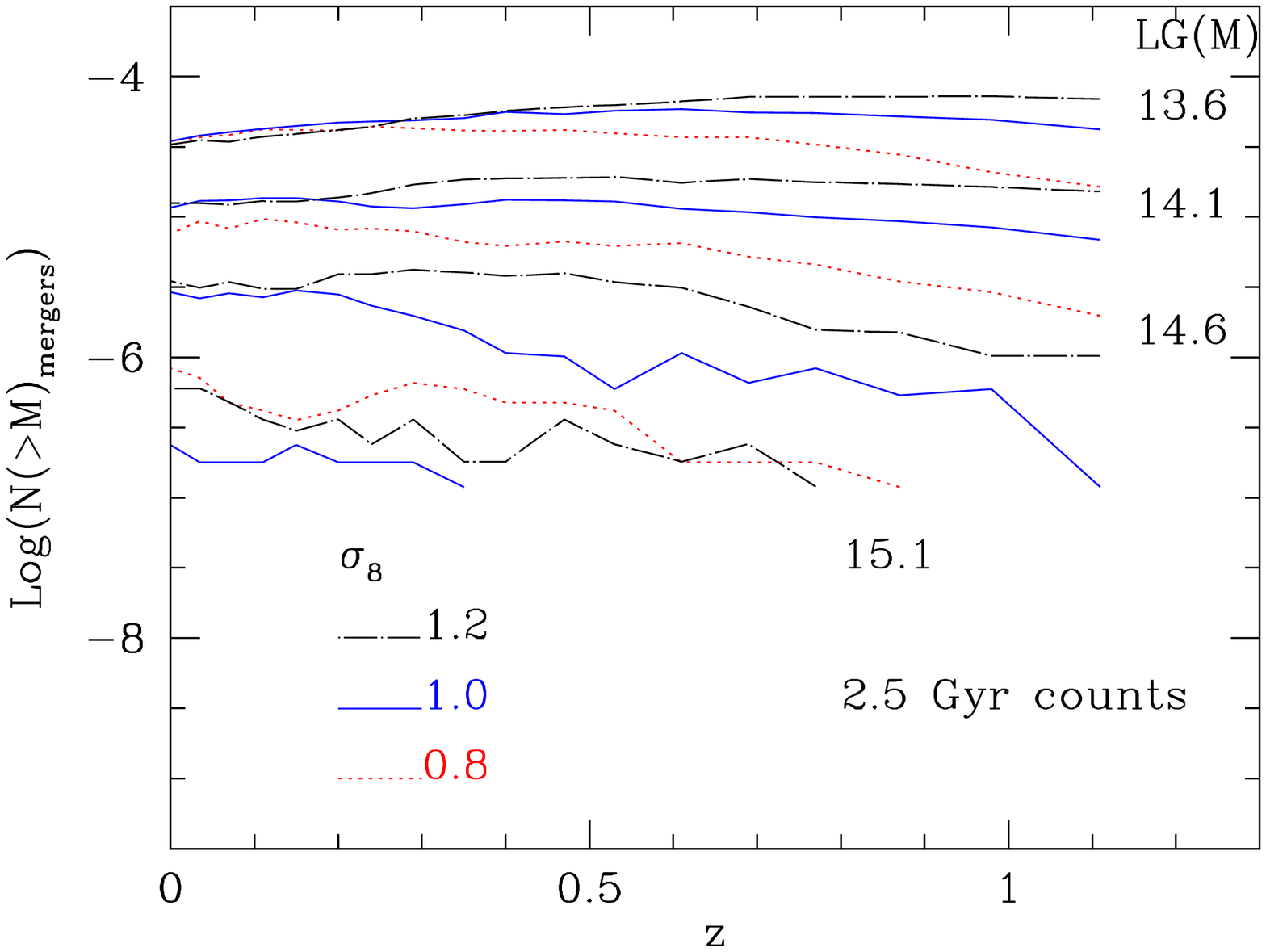}
\end{center}
\caption{Simulated major merger counts as a function of redshift for 4
different masses.  The number of recent mergers increase in time and
then decrease, perhaps due to the fewer numbers of halos at earlier
times.  The time period is $0.5$ Gyrs for the left panel and
$2.5$ Gyrs for the right panel.}
\label{fig:nms_z_sim}
\end{figure}

A related quantity, the merger counts\footnote{The merger rate, rather than
counts, as a function of redshift was studied by e.g.~Percival \& Miller
\shortcite{permil}, Percival, Miller \& Ballinger \shortcite{pmb}, along
with consequences such as star formation.} as a function of redshift at fixed
mass, is plotted in Fig.~\ref{fig:nms_z_sim}.
It appears that, for lower mass objects, the number of merger counts within
$0.5$ Gyrs first increases with look-back time and then decreases.
For higher mass objects the number merely decreases, which may be due to the
lack of available high mass objects at earlier times.
The trend is similar for the $2.5$ Gyrs merger counts, however for
$\sigma_8=1.2$ and low mass ($10^{13.6} h^{-1} M_\odot$) it appears that
there is not yet any evidence for a high redshift decline in counts.

In the next section some analytic estimates of number of major mergers
within a given interval will be made using (extended) PS theory. 

\section{Analytic estimates} \label{sec:analytic}

There has been much theoretical work on major mergers and on mergers more
generally. 
The most popular, and successful, semi-analytic formalism for predicting
merger rates and associated quantities is PS theory and its extensions
\cite{bondetal,myers,bower,lc93,lc94,kauff,kit-sut,planttree,tormen,shethone,shethtwo,sh-lem},
although there have also been merger calculations done in other frameworks
\cite{carlberg,blain,cava,cava1}.
Extended PS provides the necessary ingredients for calculating the quantity
of interest here: the number of halos of a given mass which have had a major
merger within some given time frame at some given epoch, which is
to be compared with
the simulations described above.

Press-Schechter and extended PS theory provide extremely useful
characterizations of the distributions of mass and the growth of structure.
They are computationally inexpensive and exhibit clearly the dependence
on cosmological parameters.
Their use has been widespread because, for many quantities
(e.g.~in Bond et al.~\shortcite{bondetal}, Lacey \& Cole
\shortcite{lc93,lc94}, Kitayama and Suto \shortcite{kit-sut,kit-sut2},
Tormen \shortcite{tormen},
Somerville et al.~\shortcite{sometal}), they are
in good agreement with numerical simulations.

Given the simplistic assumptions going into the PS theory, it is intriguing
that the agreement with numerical simulations is so good.
Many of the assumptions are known to be incorrect in detail:
spherical collapse \cite{smt}, the monotonic growth of halos
(e.g.~Tormen \shortcite{tormen}),
and the association of initial density peaks with final halos
\cite{carlberg,katz,frenk}.
Hence, while PS is very useful and captures something essential about the
process of structure formation, its failures \cite{gross,st,jenkins} 
can tell us something very interesting as well.  
For example, extending the picture of spherical collapse to ellipsoidal
collapse improves agreement with the mass function from numerical
simulations \cite{mona,monb,leesh,smt}.
It is to be hoped that finding where and how extended PS theory fails to
predict mergers will shed further light on this issue.

We briefly review some background from the theory of PS and extended PS
which we shall make use of below.
More detailed definitions are in the Appendix.

\subsection{Comoving number density of spherical collapsed systems}

The basic prediction of PS theory is the comoving number density of
virialized halos with mass in the range $(M, M+dM )$ at time $t$.
The PS prediction is
\eq
\label{nps}
  N_{PS}(M,t) dM =
  \frac{1}{\sqrt{2\pi}} \frac{\rho_0}{M}
  \frac{\delta_c(t)}{\sigma^3(M)} \left|\frac{d \sigma^2(M)}{dM}\right|
  \exp\left[-\frac{\delta_c^2(t)}{2 \sigma^2(M)} \right] dM \; .
\en
Here $\rho_0$ is the comoving, mean mass density of the universe,
for matter density $\Omega_m$, $\delta_c(t)$ is the threshold density contrast
for spherical collapse at time $t$ and $\sigma^2(M)$ is the variance of the
matter fluctuations smoothed over a region of radius $R$ corresponding to
mass $M=4\pi R^3\rho_0/3$.
For the functional form of $\delta_c(t)$ and a fit to $\sigma^2(R)$ for the
model under consideration see the Appendix.

In order to compare the simulations to analytic PS estimates, $\delta_c$ must
be specified.  The spherical top-hat collapse model \cite{peacock,LidLyt}
predicts that $\delta_c\simeq 1.69$ with a small cosmology dependence.
We choose to regard $\delta_c$ as a free parameter in the PS theory and
adjust it to get the best agreement with the simulations for a given choice
of group finder.
The groups found by HOP resulted in a set of halo mass distributions fit
within 40 per cent (better than 20 per cent most of the time) by PS predictions
for $M>10^{13}h^{-1}M_\odot$ if $\delta_c=1.48$ was taken.
For FOF the best fit is obtained with $\delta_c=1.64$, closer to the spherical
top-hat prediction.

\subsection{Conditional probabilities}

The random walk model inherent in the PS theory can be extended to describe
conditional probabilities: given a point in space that ends up in a halo of
mass  $M_2>M_1$ at $t_2>t_1$, the probability that it was in a halo of mass
$M_1$ at $t_1$ is:
\eq
  P_1(M_1,t_1 | M_2, t_2) d M_1 =
  \frac{1}{\sqrt{2 \pi}} \frac{\delta_{c1} - \delta_{c2}}
  {(\sigma_1^2 - \sigma_2^2)^{3/2}}
  \left\vert \frac{d \sigma_1^2}{d M_1} \right\vert
  \exp\left[-\frac{(\delta_{c1} - \delta_{c2})^2} 
  {2 (\sigma_1^2 - \sigma_2^2)} \right] d M_1
\label{p1prob}
\en
where $\delta_{c2} = \delta_c(t_2)$, $\sigma_2 = \sigma(M_2)$, etc.
Given that a point starts in a halo of mass $M_1$ at $t_1$, the reverse
probability that it ends up in a halo of mass $M_2$ at $t_2$ is
\eq
 P_2(M_2,t_2| M_1,t_1) d M_2 =
 \frac{1}{\sqrt{2 \pi}}
 \frac{\delta_{c2}(\delta_{c1} - \delta_{c2})}{\delta_{c1}}
 \left[ \frac{\sigma_1^2}{\sigma_2^2 (\sigma_1^2 - \sigma_2^2)}\right]^{3/2}
 \left\vert \frac{d \sigma_2^2}{d M_2} \right\vert
 \exp\left[ - \frac{\sigma_2^2 \delta_{c1} - \sigma_1^2 \delta_{c2}^2}
 {2 \sigma_1^2 \sigma_2^2 (\sigma_1^2 - \sigma_2^2) } \right] dM_2
\label{p2prob}
\en

\subsection{Merger rates}

Merger rates are calculated as a limiting case of the
conditional probabilities above.  The probability of a mass change
over a small time is 
\eq
P_{1,2}(M + \Delta M, t \mp \Delta t| M,t) \sim 
\frac{dP_{1,2}(M \rightarrow M+ \Delta M; t)}{dt} \Delta t
\en
if $\Delta t$ is small enough.
(Note that $P_{1,2}(M + \Delta M,t| M,t)=0$.) 
One might think of a large mass change in a small time as coming from
the addition of one halo, i.e. a merger (Lacey \& Cole \shortcite{lc93}).
With this interpretation one gets the 
\begin{itemize}
\item Rate at which a point in a halo of mass $M_1$ is 
incorporated into a halo
of mass $M_2$ at $t$:
\eq
\begin{array}{ll}
\frac{ d P_1(M_1 \to  M_2; t)}{dt} d M_1 &=  \lim_{\Delta t \to 0}
\frac{  P_1(M_1,t - \Delta t | M_2, t)}{ \Delta t} d M_1 
\\ &=
\frac{1}{\sqrt{2 \pi}}
\frac{1}
{(\sigma_1^2 - \sigma_2^2)^{3/2}}\left[-\frac{d \delta_c(t)}{dt}\right]
\left\vert \frac{d \sigma_1^2}{d M_1} \right\vert d M_1 \; .
\end{array}
\en
\item
Rate at which a halo of mass $M_2$ is formed from addition of mass to
a halo of mass $M_1$ at $t$, per point:  
\eq
\begin{array}{l}
\frac{d P_2(M_1 \to M_2;t)}{dt} d M_2  = 
\lim_{\Delta t \to 0} \frac{d P_2(M_2, t+ \Delta t| M_1 t)}{\Delta t} d M_2
\\  \phantom{xxxxx} = 
\frac{1}{\sqrt{2 \pi}}
\left[ \frac{\sigma_1^2}{\sigma_2^2 (\sigma_1^2 - \sigma_2^2)}
\right]^{3/2} \left[-\frac{d \delta_c(t)}{dt}\right]
\left\vert \frac{d \sigma_2^2}{d M_2} \right\vert 
\exp \left[
- \delta_{c}(t)^2  \left(\frac{1}{2 \sigma^2_2}  -
\frac{1}{2 \sigma^2_1}\right) \right] d M_2
\end{array}
\en
\end{itemize}

These bulk probabilities can be combined to calculate the number of
halos of a given final mass that have undergone a major merger
(i.e.~have predecessors with a given ratio) within a certain time
interval, with the latter chosen to correspond to the relaxation
process of interest.  (We note that
a modified version of the
PS formalism that differentiates between (instantaneous)
major mergers and accretion has been developed in \cite{raig,salvador}
as well.)

Heuristically, one can count the major mergers which have occurred at a
specific time for halos of a given final mass, whether or not they accreted
mass before or after the merger, by taking the 
\eq
\begin{array}{cl}
&{\rm starting \; \# \; of \; halos}\\
\times&{\rm merger \; prob. \; per \; unit  \; time}\\
\times&{\rm prob. \; to \; end \; up \; in\; final \; mass \; range} \; .
\end{array}
\en
This is for a specific time and must be integrated to study a time
interval.  Plugging in the probabilities and integrating over time
this becomes 
\eq 
\label{mercalc}
\begin{array}{ll}
N_{{\rm mer}_{direct}}(M_2, t_1,t_2)dM_2 =& \int_{t_1}^{t_2} 
dt \int d(M_1 + \Delta M_1) \int dM_1  N_{PS}(M_1,t) \\
& \times \frac{d P_2}{dt}(M_1 \rightarrow M_1 + \Delta M_1;t) 
P_2(M_2,t_2| M_1 + \Delta M_1, t) \; .
\end{array}
\en
This will be referred to as the ``direct'' estimate.\footnote{After 
the bulk of the work in this note was completed, the authors became
aware of a similar expression in \cite{cava2}.} 
In words, this is the probability that at some time $t$ in a given time
interval of size $t_2-t_1$ a starting halo mass of mass $M_1$ increases
its mass instantaneously by $\Delta M_1$, times the probability that the
resultant $M_1 + \Delta M_1$ mass ends up, at $t_2$, in a halo of mass
of $M_2$.  Taking $M_1$ to be the mass of the larger of the
predecessors, the range of integration for $\Delta M_1$ (the mass of
the second largest predecessor) is determined by the chosen merger
mass ratio.
This allows for the possibility of a merger happening any time between
$t_1$ and $t_2$, with accretion before or
afterwards, which is in some sense the key difference between this and
the work of Carlberg \shortcite{carlberg} as formulated by Lacey \& Cole
\shortcite{lc93}.  Another differences is the range of integration for
the masses taken.  An equivalent way of writing the above\footnote{The
factor $M_2/M_1$ is due to the one to one correspondence (by assumption)
between initial halos of mass $M_1$ and final number of  halos.
If a halo of mass $M_1$ ends up in a halo of mass $M_2$ then all the mass
in the $M_1$ halo ends up in the halo of mass $M_2$.  So one can count the
number of initial, and thus final, halos by taking the mass fraction and
dividing by $M_1$ to get the corresponding number of halos.} is
\eq
\int dt \int d(M_1 + \Delta M_1) \int dM_1 
\frac
{d P_1}{dt}(M_1 \rightarrow M_1 + \Delta M_1;t) P_1(M_1 + \Delta M_1,
t| M_2, t_2) N_{PS}(M_2, t_2) \frac{M_2}{M_1} 
\en

The limits on the integrals are as follows.
The time integral is taken to correspond to the time interval of interest
($0.5$ or $2.5$ Gyrs), and the range of $\Delta M_1$ is determined by the
mass ratios required for a major merger.  So for our case, with ratios
$1:1$ to $1:5$, $M_1/5 \le \Delta M_1 \le $Min$(M_f-M_1, M_1)$.
For the $dM_1$ integral, the chosen mass ratio fixes the upper limit to
be $5 M_2/6$.  The lower limit is not obvious, but it
turns out that the dependence is rather weak.  By considering the
simulations one finds that $M_1 \in [M_2/3, 5 M_2/6]$ will cover most
of the range of largest predecessor masses for 0.5 Gyrs. This was also
used for the 2.5 Gyr run.  To use this formula more generally one
would like to be able to choose this lower limit without relying upon
simulations.  Changing this lower limit for the
0.5 Gyr case between $M_2/2.5$ and $M_2/3.5$ produced indistinguishable
results.\footnote{We thank
the referee for suggesting we calculate this dependence.}
  To evaluate Eq. (\ref{mercalc}), two of the three
integrals can be done exactly, but the last integral over $M_1$ seems
to require numerical integration (see Appendix).
A quadratic fit to the corresponding $\log
N_{\rm mer}(>M)$, for $\delta_c = 1.48$ as a function of $\log(M)$ and
$y = \ln(1+z)$ is shown as the dashed line in Figs.~\ref{fig:hz0}
and \ref{fig:hz1}.

It should be noted that even within the PS description
this expression is only approximate.  If a halo has two jumps before
getting to the final mass and they are both a sizeable fraction if its
mass, they may be counted twice.  In principle one could bound this
effect by calculating the probability of two jumps occurring
using a generalization of Eq.~(\ref{mercalc}). 
One will also double count the number of mergers where the two initial
masses are identical, but this is expected to be a small
number.  Note that the distribution of mergers with a given mass
change should be independent of the initial mass, using the random
walk interpretation of the PS formulae 
\cite{percetal}.  However, for a {\it major} merger the given mass change
of interest depends on the mass of the initial halo, and in addition
the total number of mergers depends on the number of initial halos
(which depends on both time and mass), consequently some dependence on
mass and era is expected. 

There are other approximate expressions which can be considered.
Here we focus on one which has a particularly simple origin, in order to see
how well it captures general trends and the quantity of interest.
This is to use a jump to $M$ instantaneously as an estimate of merger counts.
Saying that the largest mass component of the two halo merger has at least
half the final mass and allowing it to jump by a mass in the range
($M/2, 5 M/6$) corresponds to taking
\eq
\begin{array}{ll}
\label{psmer}
N_{{\rm mer}_{jump}}(M,t) 
\equiv &\int_{\frac{1}{2}M}^{\frac{5}{6}M} 
d M_1 \frac{d P_1(M_1 \to M;t)}{dt} \frac{M}{M_1}N_{PS}(M,t)
\end{array}
\en
This will be referred to as the ``jump'' estimate.
Requiring the largest predecessor to have mass only $\ge M/2$ avoids counting
changes in both components of a major merger as distinct mergers (except in
the case, probably rare, where both initial masses are exactly equal to half
the final mass).  However, there  is some under-counting expected because, as
noted earlier, the simulations showed that the largest predecessor
could sometimes be as little as 1/3 the final mass.  This expression only
counts the masses which come exactly to a given mass at the given
time, rather than any masses which might have merged and then later
accreted to reach the given mass, or those which have reached this
mass and then accreted out of the mass range.   As the time dependence
factors out in front, the only difference between $2.5$ Gyrs and
$0.5$ Gyrs is the multiplication by an overall factor of $5$ to change
time units. 

This turns out to be a simple quantity since for the range of
interest the effect of the factor $M/M_1$ in Eq.~(\ref{psmer})
gives just an overall prefactor of approximately 1.44:
\eq
\begin{array}{ll}
 N_{{\rm mer}_{jump}}(M,t)  &\simeq  
1.44 \int_{\frac{1}{2}M}^{\frac{5}{6}M} 
d M_1 \frac{d P_1(M_1 \to M;t)}{dt}N_{PS}(M,t)
\\ &=
1.44 
\sqrt{\frac{2}{\pi}}\left( \frac{1}{\sqrt{\sigma^2(\frac{5}{6}M) - 
\sigma^2(M)}}-
 \frac{1}{\sqrt{\sigma^2(\frac{1}{2}M) - \sigma^2(M)}}\right)
\left[-\frac{d \delta_c(t)}{dt}\right] N_{PS}(M,t) \; .
\end{array}
\en

Another approximation which is sometimes used as the formation rate
(e.g.~Kitayama and Suto \shortcite{kit-sut})\footnote{There is no factor
of $M/M_1$ here because the number of initial halos is not in one to one
correspondence with the number of final halos.  For example, a huge number
of infinitesimally sized halos can go to form one final halo.},
\eq
\label{ksform}
\begin{array}{ll}
N_{{\rm mer}_{form}}(M,t;M/2) 
& \equiv\int_0^{M/2} d M_1 \frac{d P_1(M_1 \to M;t)}{dt} N_{PS}(M,t) \\
& = \sqrt{\frac{2}{\pi}} \frac{1}{\sqrt{\sigma^2(M/2) - \sigma^2(M)}}
\left[-\frac{d \delta_c(t)}{dt}\right] N_{PS}(M,t) \; .
\end{array}
\en 
which turns out (as seen by numerical integration in the regime of interest)
to also be close to the ``jump'' estimate when multiplied by 1.44:
\eq
N_{{\rm mer}_{jump}} \sim 1.44 N_{{\rm mer}_{form}} \; .
\en

The ``jump'' estimate is somewhat larger, by about 4 per cent, for masses
$\sim 5 \times 10^{13} h^{-1} M_\odot$ and $\sim$9 per cent larger for masses
$\sim 1.5 \times 10^{15} h^{-1} M_\odot$.
In both of these, one is only looking at mergers going up to the final
mass of interest, while in reality accretion should be taking halos
both in and out of the range of interest.  These quantities will turn
out to work surprisingly well, perhaps because of some averaging out
from the accretion effects.

There has been related analytic work using extended PS to calculate
``formation times.''  These differ from the merger counts under study here.
A few differing formation time definitions are used,
e.g. when half the mass in the halo has
the halo has assembled \cite{lc93,kit-sut} or when the assembly of mass in
the halo ends \cite{blain,sasaki,permil}.  For example,
a major merger by our definition could occur after half of the halo mass 
assembled, e.g. with
a halo of half the final mass merging with a 
halo of one third the final mass.    
Analytic and numerical formation times have been compared,
e.g.~by Tormen \shortcite{tormen}.

\section{Comparisons} \label{sec:compare}

The analytic formulae in section \S3 for the number of mergers, the
``direct'' and ``jump'' calculations (Eqs.~\ref{mercalc}, \ref{psmer})
in the previous section, were compared with the simulation results described
in section \S2 for mass starting at $5 \times 10^{13} h^{-1}M_\odot$ and above.
As mentioned earlier this minimum mass was chosen because a major
merger could have a smaller predecessor of mass $1/10$ the final mass,
and so the minimum final mass was taken to be $10$ times the minimum
halo mass.  The two analytic estimates differed from each other and
consequently had different success in fitting the simulation results.
The goodness of fit also depended on the time interval ($0.5$ or
$2.5$ Gyrs) considered and (very weakly) on the group finder used.

Comparison with Eq.~(\ref{mercalc}), the ``direct'' estimate,
will be made first.   The analytic expressions were integrated numerically
and then fit quadratically (in $\log(M)$ and $\ln(1+z)$) and these quadratic
fits were compared with the simulations.   A quadratic best fit to the
simulations directly, for all masses and redshifts, could vary from
the binned simulation values by up to $40$ per cent in the $0.5$ Gyr runs --
thus exact agreement with any smooth  curve is not expected to be better than
this.
For $0.5$ Gyrs and using HOP, the quadratic fit to the ``direct'' formula
was in broad agreement with that found in the simulations.
The analytic calculation tended to over-predict the number of low mass
($M < 10^{14} h^{-1} M_\odot$) halos which had major mergers, and this
occurred more often at early times and lower $\sigma_8$, being within a
factor of three in the worst case and within $40$ per cent much of the time.  
This may be in part due to the known tendency of PS to
overestimate the number of halos, and thus predecessors
for major mergers, at low mass (Gross et al \shortcite{gross},
Somerville et al.~\shortcite{sometal}).
A competing effect is the ``loss'' of small halos in the simulations.
Another factor may be the residual effects of the initial conditions,
again strongest at early times and smallest $\sigma_8$.
For mass $\sim 10^{14} M_\odot$ the ``direct'' prediction was close,
over-predicting for lower mass and under-predicting for higher mass.
(For example, for $\sigma_8= 0.8$ and the $0.5$~Gyr time frame, the predicted
``direct'' merger  counts for final halos of mass $M< 10^{14}$ were too high
by more than $40$ per cent down to $z \sim 1$.)   For the $2.5$~Gyr runs, the
agreement at low mass was a lot worse between the ``direct'' estimate
and the simulations, with the number of mergers predicted for mass
less that $10^{14}h^{-1}M_\odot$ over-predicted by $40$ per cent for almost
all times and values of $\sigma_8$.
The fit improved with increasing $\sigma_8$. 

Surprisingly, the ``jump'' estimate (Eq. \ref{psmer})
worked quite well.
The ``jump'' estimate was unbiased for $0.5$~Gyr but for
$2.5$~Gyr tended to overestimate the number of mergers.
As noted, the ``jump'' estimate was
slightly larger than 1.44 times the formation rate based estimate 
and closer to the
simulations for $0.5$~Gyrs.   
Note that the ``jump'' and formation rate based estimates
have time dependence only in an overall factor out front, 
taking the time units to be the interval of interest.  In particular, 
for the two intervals chosen here, 0.5 Gyrs, and 2.5 Gyrs, 
the predictions only differ by a factor of 5.  
This scaling was not observed in the simulations, which differed by a smaller
amount.
Thus if the ``jump'' or formation rate based 
estimate works for one time, it will
not work very well for the other.  We found that it worked well for
$0.5$~Gyrs (i.e. within 40 per cent most of the time) but not for $2.5$~Gyrs
(consistently too high, centered at an overprediction of counts by 
$\sim$40 per cent).
   
\begin{figure}
\begin{center}
\leavevmode
\epsfxsize=3.2in\epsfbox{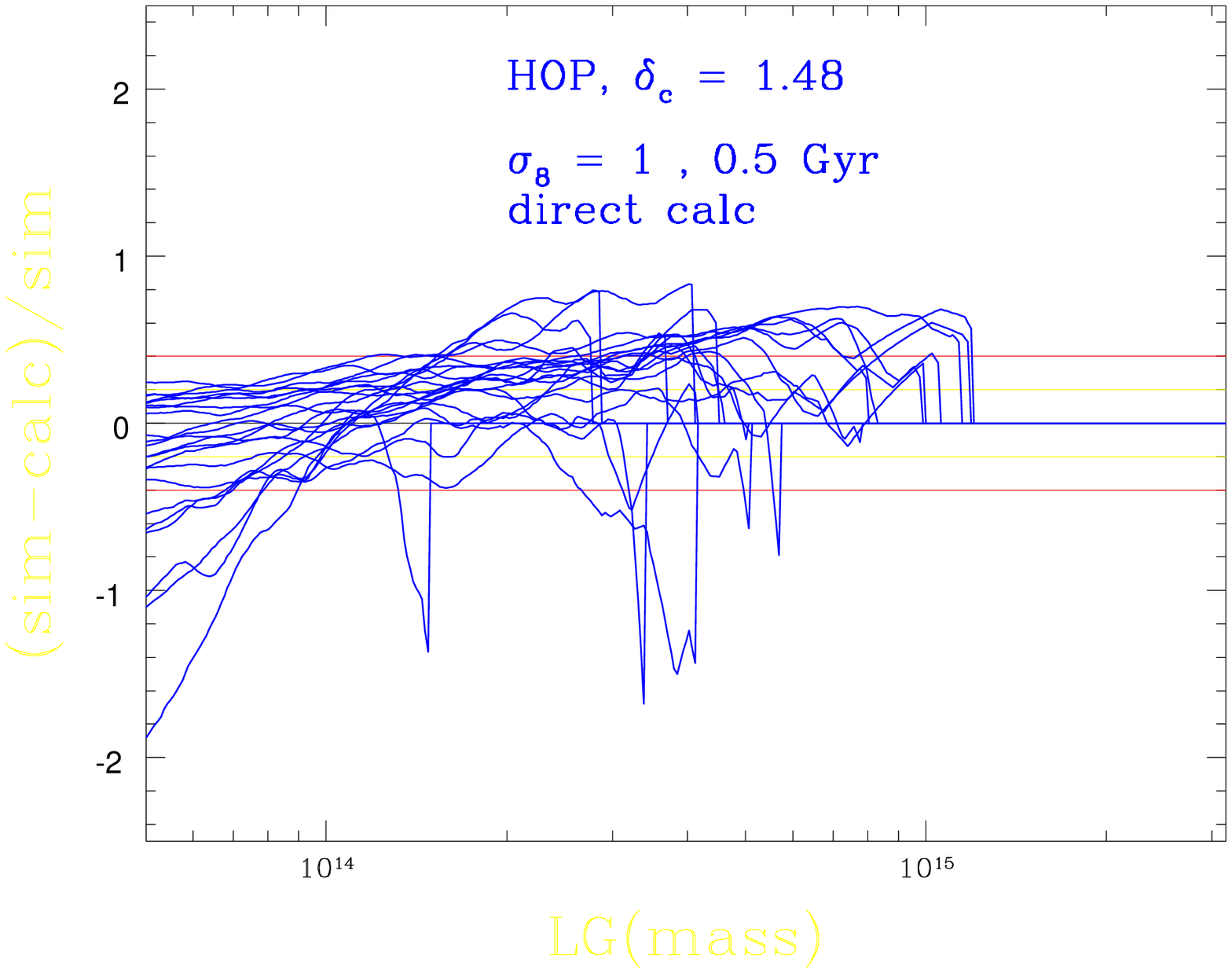}
\epsfxsize=3.2in\epsfbox{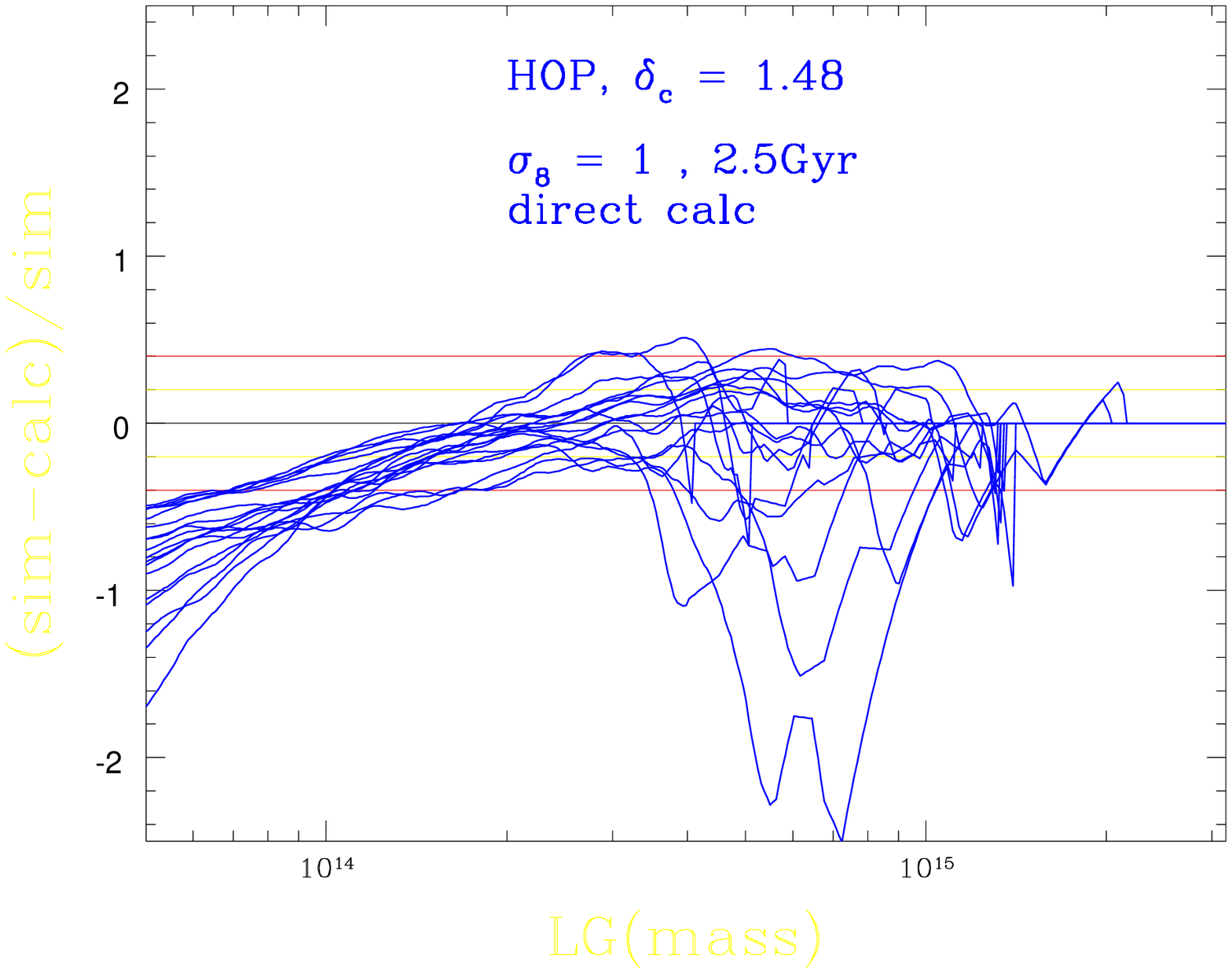}
\\
\leavevmode
\epsfxsize=3.2in\epsfbox{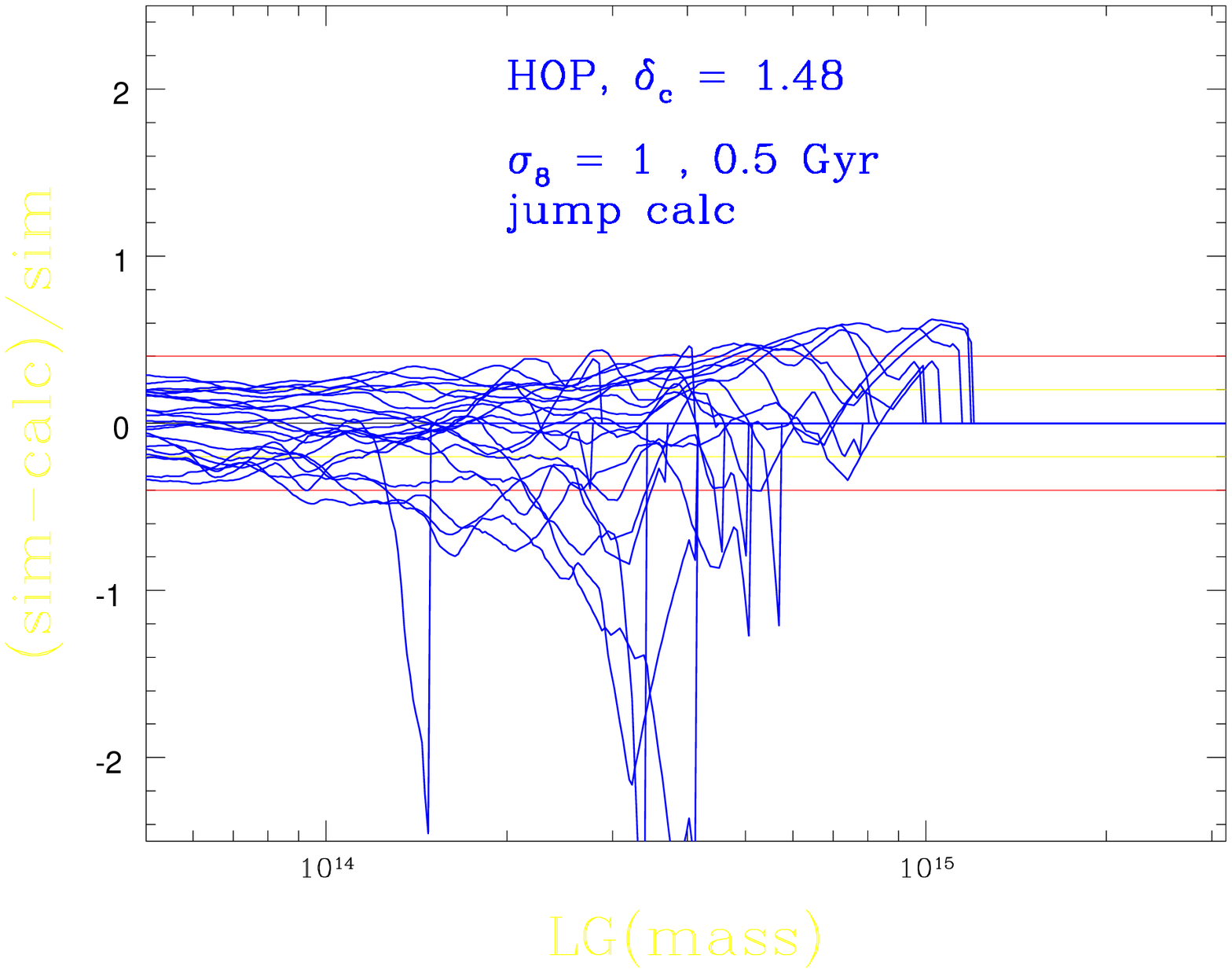}
\epsfxsize=3.2in\epsfbox{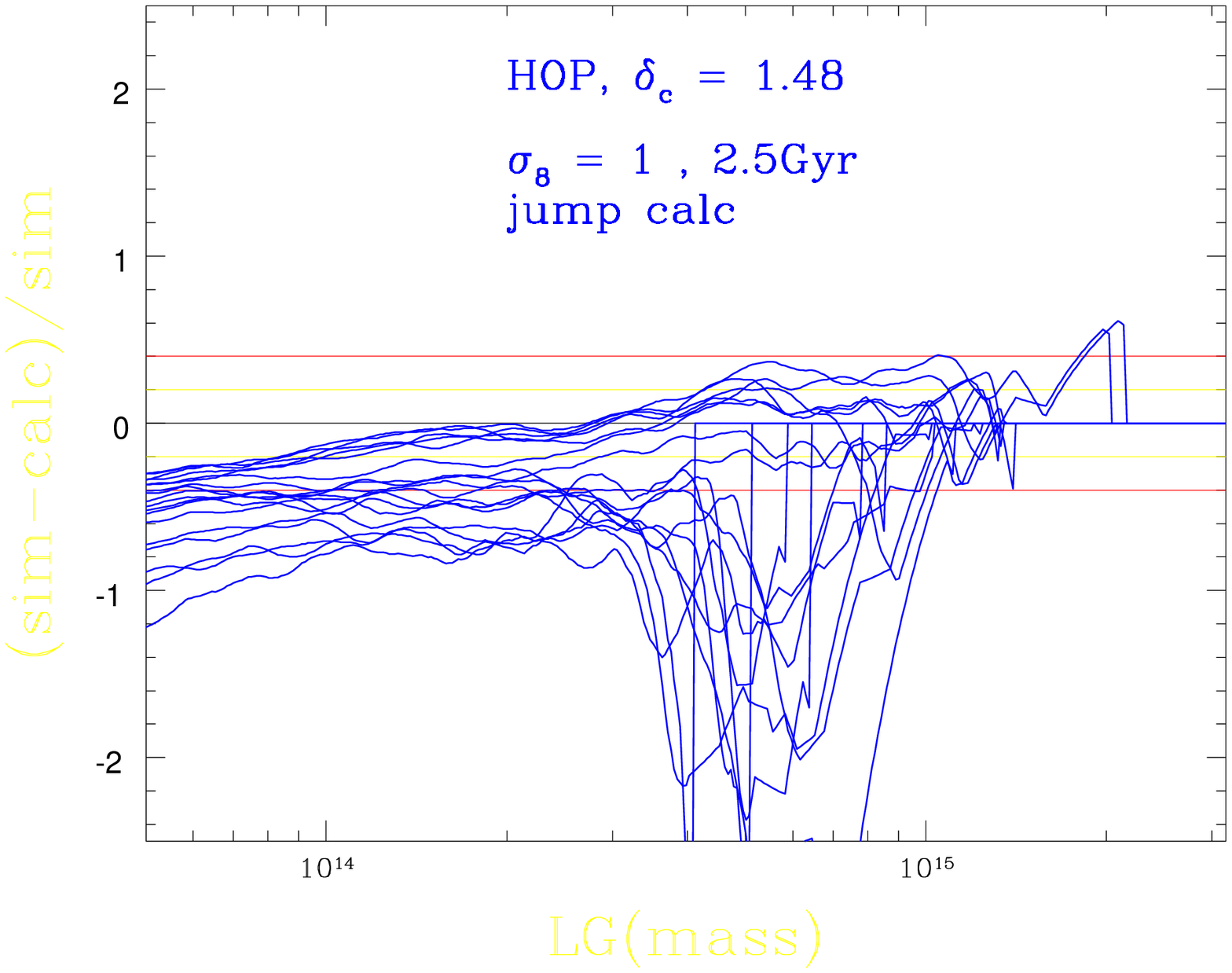}
\end{center}
\caption{``Direct'' and ``jump'' predicted merger counts
vs. simulations: Eqs. (\protect\ref{mercalc}, \protect\ref{psmer}) for $0.5$
and $2.5$~Gyrs.  The horizontal lines indicate $0$, $20$ and $40$ per cent
deviations between the simulation and the analytic estimate.  The
simulation counts have been smoothed over a range of $.05$ in $\log M$
before comparing with the analytic expressions.}
\label{fig:ctssims}
\end{figure}

The percentage deviation for the two time intervals and the ``direct''
and ``jump'' estimates, for $\sigma_8 = 1$, is shown in Fig.~\ref{fig:ctssims}.
Each line corresponds to a different time step, and horizontal lines
mark $0$, $20$ and $40$ per cent deviation from $0$.  Large spikes at
high mass are often due to small number statistics (which are worse at early
times).
For the ``jump'' calculation, the trends for the fits
(within 40 per cent for 0.5 Gyrs and centered at 40 per cent too high
for 2.5 Gyrs) were the similar for the other values of $\sigma_8$.

In Fig. 6,
the RMS deviation of the calculated versus simulated counts and
the average deviation are shown as a function of redshift
for all three values of $\sigma_8$
for both the ``direct'' and ``jump'' predictions.

\begin{figure}
\begin{center}
\leavevmode
\epsfxsize=3.2in\epsfbox{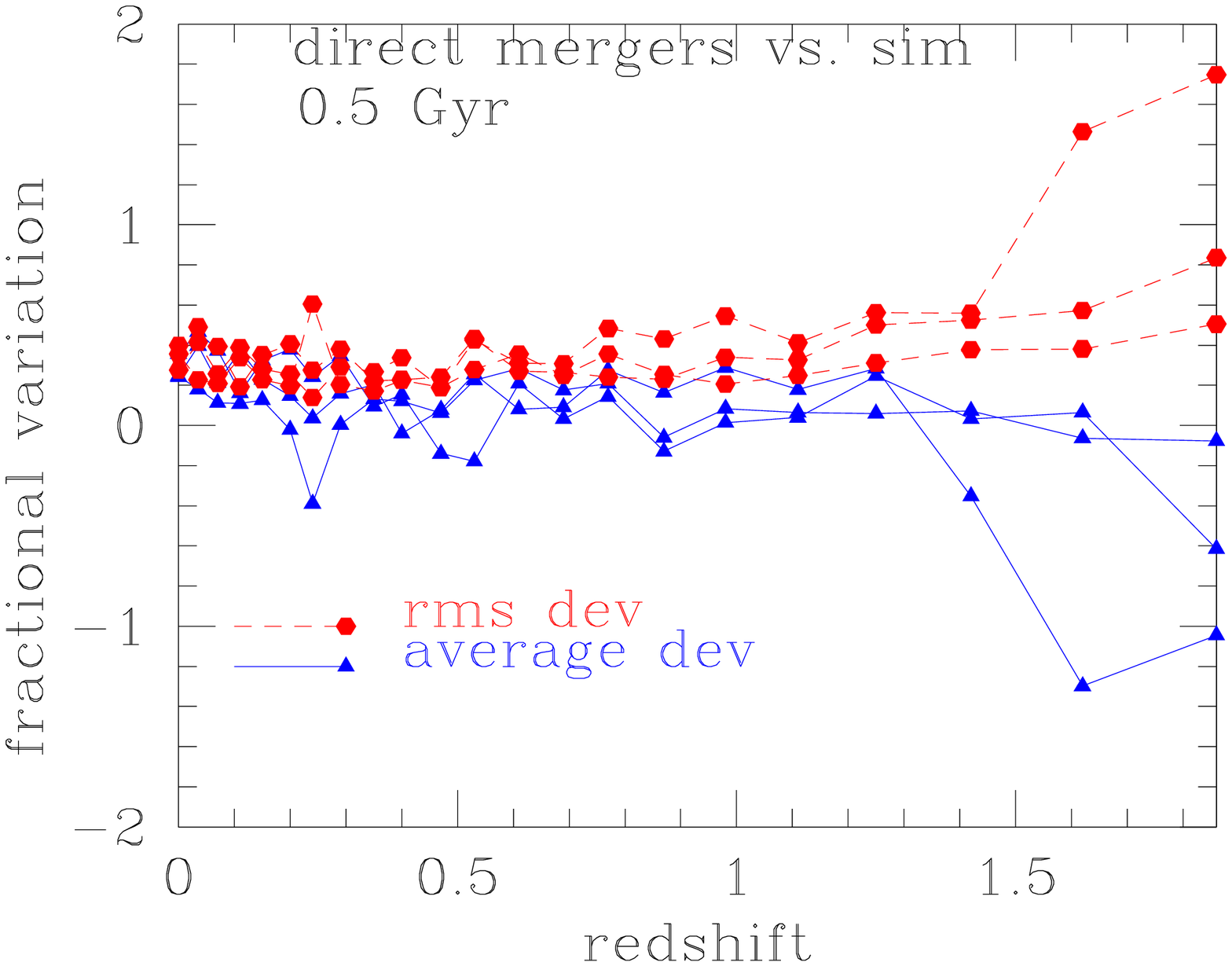}
\epsfxsize=3.2in\epsfbox{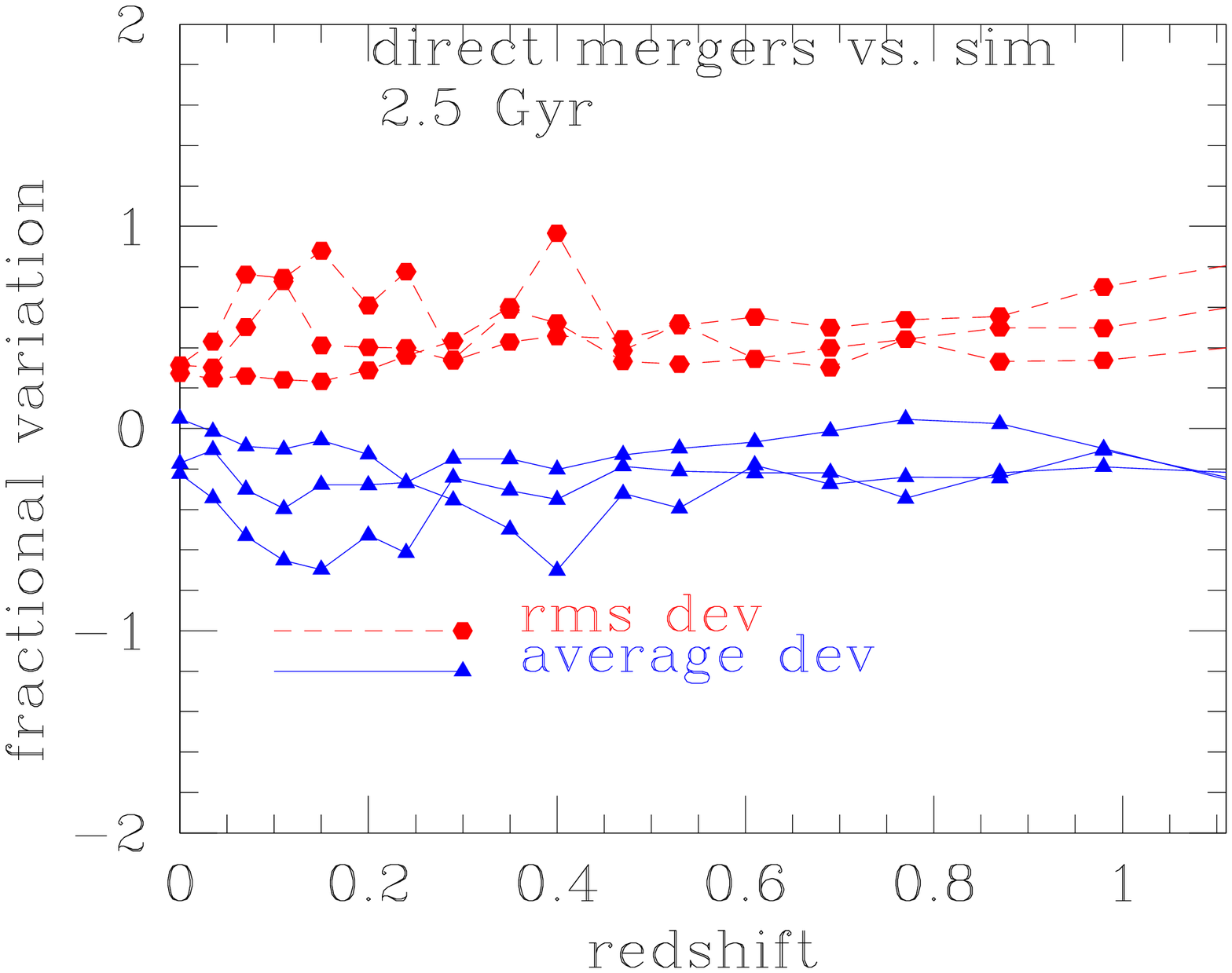}
\\
\leavevmode
\epsfxsize=3.2in\epsfbox{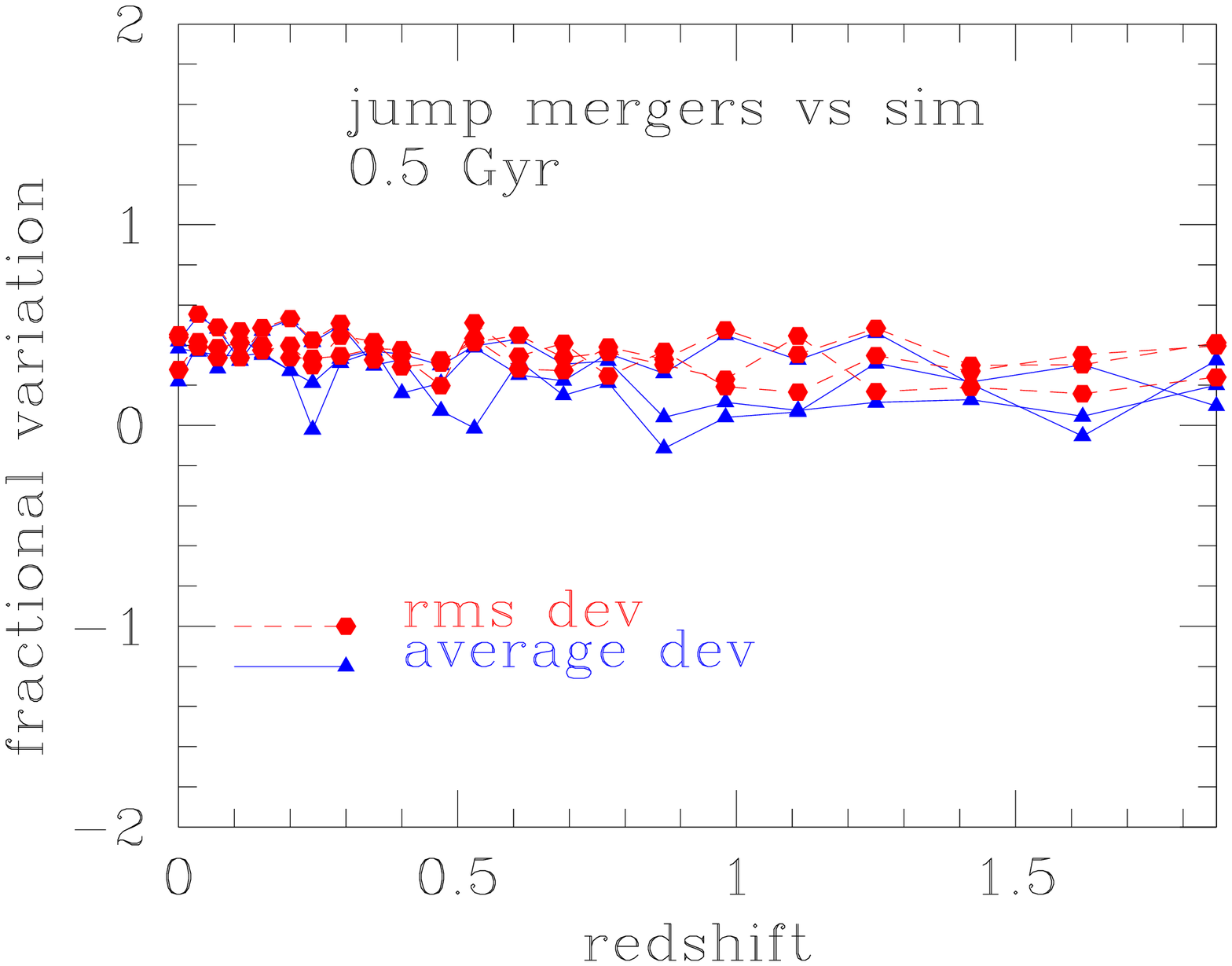}
\epsfxsize=3.2in\epsfbox{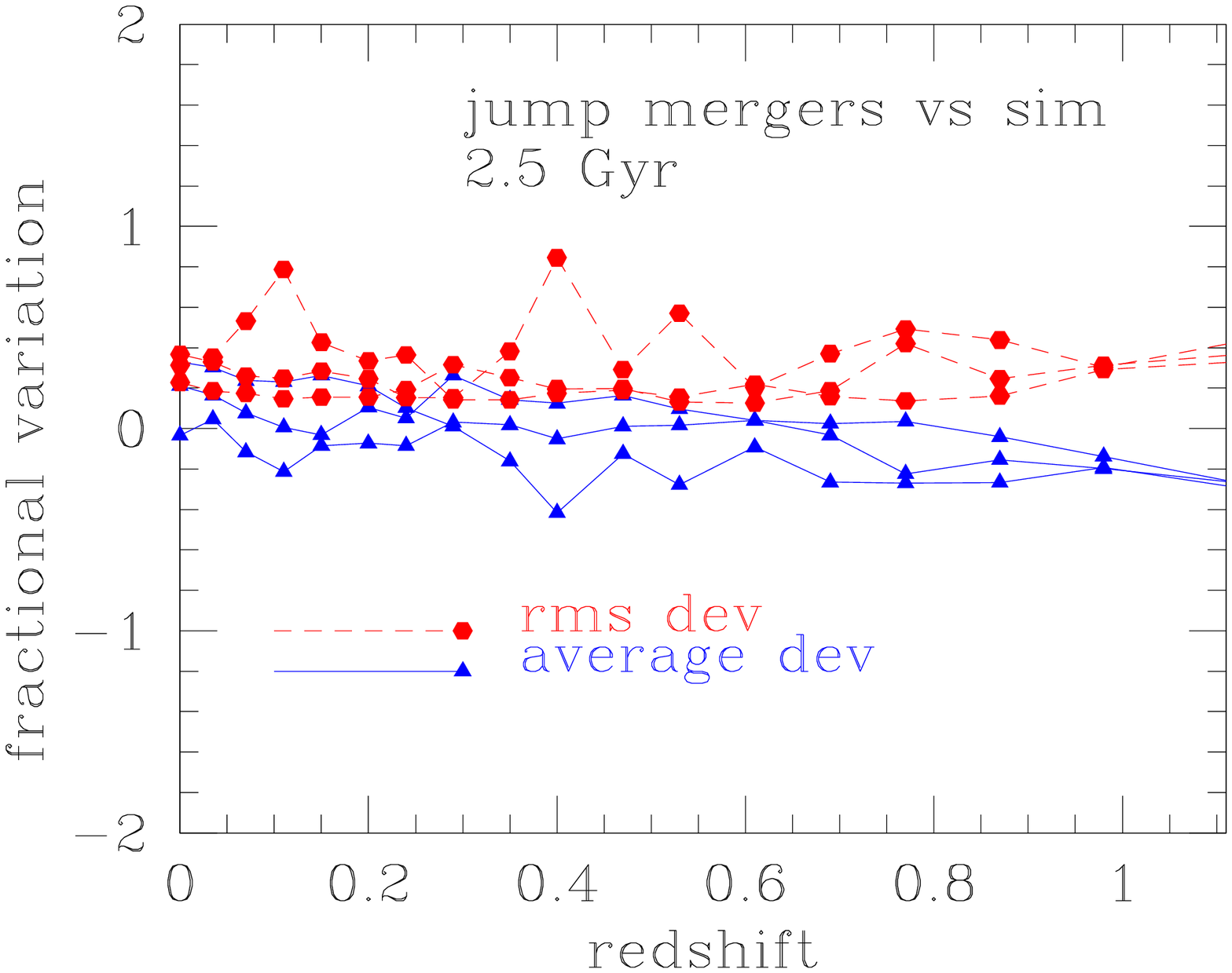}
\end{center}
\caption{RMS and average deviation as
a function of redshift for $\sigma_8 = 0.8, 1.0, 1.2$
for ``direct'' (top) and ``jump'' (bottom) fits,
Eqs. (\protect\ref{mercalc}, \protect\ref{ksform}).} 
\label{fig:devs}
\end{figure}

Comparing the HOP and FOF group finders for
the $\sigma_8=1$ case, some notable differences for
merger counts and number counts were noted, as mentioned earlier.
However, the analytic estimates worked similarly for both
group finders when the appropriate $\delta_c$ was used.

In conclusion, it seemed that both analytic estimates gave comparable
numbers of mergers to what was found in the simulations.
The ``jump'' estimate
worked better than the ``direct'' estimate in matching the
shape of the curve $N_{\rm mer}(>M)$, but over-predicts counts for all
masses for longer times.  It is of clear interest to pursue this,
but doing so will require a larger mass and redshift range and
larger simulations.

It is also interesting to try to find some estimate of the
merger fraction from the simulations alone, similar to
``universal'' fitting functions for numbers of
halos as functions of various parameter combinations which have been found
to have a particularly simple form (e.g. Sheth-Tormen
\shortcite{st}, Jenkins \shortcite{jenkins}). We found that, using
$\nu
= \delta_c/\sigma(M,z)$\footnote{In 
terms of the often used
parameter $M^*$, $M = M^* \leftrightarrow \nu = 1$.},  
the combinations
$
\frac{N_{\rm mer} (>M)}{N(>M)} \sim \nu(1+z)^{1.5(1)}$
for 0.5(2.5) Gyrs and 
$\frac{N_{\rm mer}(M)}{N(M)} \sim \nu(1+z)^{1.5}$
seem to be promising possible scalings.  As
the relatively small number of mergers makes these quantities very
noisy, bigger simulations and better statistics will allow
further study of this question.

\section{Conclusions}

In this note we have concentrated on the question of how many halos of
a given mass and time have had a major merger (ratio of two
largest predecessors ranging from $1:1$ to
$1:5$) within the last $0.5$ and $2.5$~Gyrs, for the mass range $5
\times 10^{13} h^{-1} M_\odot \sim 10^{15} h^{-1} M_\odot$.  Numerical
simulations for $3$ values of $\sigma_8$ in a $\Lambda$CDM cosmology
were compared with two analytical estimates.  These counts can be related to
observational processes with $0.5$~Gyr and $2.5$~Gyr relaxation
times.  For $0.5$~Gyr,
the ``jump'' analytic estimate, Eq. (\ref{psmer}), was
closest to the simulations, although the ``direct'' estimate,
Eq.~(\ref{mercalc}), was close for low enough redshift and high enough
$\sigma_8$. 
Ratios of major merger counts were also considered.
The main conclusion is that
the PS estimates of merger counts,
a very interesting quantity from the observational point of view,
roughly agree with simulations.  This suggests that
future study of this approach may be fruitful.  Many extensions and
refinements are possible, such as using
larger simulations, going back to higher
redshifts, and varying parameters such as time
intervals and mass ratios.  The dynamical disruption caused by the
major merger may be further characterized by including additional
information about the predecessors and daughter halos.  For instance, 
our criteria do not discriminate between the transfer of
a bound subhalo or the stripping of outer (perhaps spatially very
far apart and dynamically unrelated) particles of a parent halo to a daughter
halo.  The former would be expected to be more likely to
cause more dynamical 
disruption.\footnote{We thank the anonymous referee for suggesting
we introduce this distinction.}
In addition, the analytic calculations 
may be able to use the more accurate fitting functions for masses
(Sheth and Tormen \shortcite{st}, Jenkins et al
\shortcite{jenkins}), if these approaches can be generalized
to extended PS conditional probabilities.

JDC thanks J. Mohr, R. Sheth, and R. Somerville for discussions.
JDC was supported in part by NSF-AST-0074728.
During his stay at the Center for Astrophysics JSB was supported by
NSF-PHY 95-07695.
M.~White was supported by the US National Science Foundation and a
Sloan Fellowship.
JDC and MW thank the ITP for hospitality during the completion of this
work and and during that visit were supported in part by the
National Science Foundation under Grant No. PHY94-07194.  
Parts of this work were done on the Origin2000 system at the National
Center for Supercomputing Applications, University of Illinois,
Urbana-Champaign.
We thank the referee for many helpful comments and questions.

\appendix

\section{Press-Schechter and extended Press-Schechter}

\subsection{Notation}

In numerical simulations it appears that many properties of the final
density field are present in the initial conditions and are simply
``sharpened'' by the non-linear amplification of gravity.
Press Schechter \shortcite{ps} theory utilizes this in an essential way,
and couples it with the theory of non-linear evolution of a uniform,
spherical overdensity embedded in a homogeneous universe.
For this reason we need to be able to predict the scale- and time-dependence
of density fluctuations in linear theory.

We shall convert between mass and length scales using a top-hat filter in
real space, i.e.~the mass associated with a smoothing scale $R$ is
$M=(4\pi/3)\pi R^3\rho_0$ where $\rho_0$ is the comoving mean mass density
of the Universe,
\begin{equation}
\rho_0 = 2.7755 \times 10^{11}\ \Omega_m
  \quad (h^{-1} M_{\odot})(h^{-1} {\rm Mpc})^{-3}
 \qquad .
\end{equation}

We assume that the initial fluctuations were Gaussian with zero mean.
The variance of the fluctuations, smoothed over a region of mass M,
is given by  
\eq
  \sigma^2(M) \equiv \sigma^2(M,z=0)
  = \int_0^\infty \frac{dk}{k} \,  W^2(k R(M)) \Delta^2_\delta(k) 
\en
where $W(kR)$ is the window function for top-hat filtering
\eq
  W(kR) = 3 \left( \frac{\sin(kR)}{(kR)^3} - \frac{\cos(kR)}{(kR)^2}
\right) \; . 
\en
The power spectrum (today) is given by
\eq
  \Delta^2_\delta(k) \equiv
  {k^3P(k)\over 2\pi^2} =
  \delta_H^2(k) \frac{k^4}{H_0^4} T^2(k) 
\en
which uses the primordial density spectrum $\delta_H(k) \sim k^{n-1}$,
for $n \sim 1$.  We take $n=1$.  For the transfer function in both the
calculations and the simulations we used (Efstathiou et al.~\shortcite{ebw}) 
\eq
  T(k) = \left[ 1 + (a k + (b k)^{3/2} + (c k)^2)^\nu \right]^{-1/\nu}
\en
where 
\eq
  a = (6.4/\Gamma) h^{-1} {\rm Mpc} , \; 
  b= (3/\Gamma) h^{-1} {\rm Mpc} , \;
  c = (1.7/\Gamma) h^{-1} {\rm Mpc} , \;
  \nu = 1.13
\en
In practice $\sigma^2(M)$ was calculated up to an overall constant which is
fixed by the choice of $\sigma_8$.  A good fit near $R=8 h^{-1} {\rm Mpc}$ is 
\eq
 \log_{10}\left( {\sigma^2(R)\over\sigma_8^2}\right) =
 -1.4936 \log R_8 - 
 0.46554 (\log R_8)^2  - 
 0.0982478 (\log R_8)^3
\en
where $R_8 = \left(R/8 h^{-1} {\rm Mpc}\right)$.
This is a fit to numerical integration between $0.27 \le \log R \le 1.49$
which is better than $\sim 1$ per cent.

Collapsed halos are taken to be regions in the {\it linear\/} density
field with density greater than some critical density contrast, $\delta_c$,
when smoothed on a scale $R$.
In practice we take account of the linear growth by holding the variance
fixed and reducing the density threshold
\eq
  \delta_c(t(z)) =  (1+z) \frac{g(\Omega_0)}{g(\Omega(z))} \delta_c \; .
\en
We found $\delta_c(z=0) = 1.48$ was a good fit to the mass distribution of
the halos found by HOP \cite{hop} and $\delta_c(z=0) = 1.64$ a good fit for
those found with FOF \cite{fof}.
The growth rate can be approximated as \cite{cpt,viana}
\eq
g(\Omega (z)) = 
\frac{5}{2} \Omega \left[ \frac{1}{70} + \frac{209}{140} \Omega -
 \frac{\Omega^2}{140} + \Omega^{4/7} \right]^{-1}
\en
and 
\eq
\Omega(z) = \Omega_m \frac{(1+z)^3}{1 -\Omega_m + (1+z)^3 \Omega_m}
\en
for a flat $\Lambda$ universe.

\subsection{Formalism}

The Press-Schechter formalism has been extended to describe histories
of halos (Bond et al.~\shortcite{bondetal}, Bond \& Myers \shortcite{myers},
Bower \shortcite{bower}, 
Lacey and Cole \shortcite{lc93}, \shortcite{lc94}, Kitayama
\& Suto \shortcite{kit-sut}, Baugh et al.~\shortcite{baugh}, Somerville \&
Kolatt \shortcite{planttree},
Tormen \shortcite{tormen}).
A useful description can also found in the textbooks by
Peacock~\shortcite{peacock} and Liddle \& Lyth~\shortcite{LidLyt}.

Press-Schechter and extended Press-Schechter theory have a
heuristic random walk interpretation based on consideration of
trajectories.  Each trajectory corresponds to the
motion of a point in space under the addition of modes $\delta_k$
(i.e.~filtering the density field around each point in space with filters
of steadily increasing $k$).
The arguments are rigorous when this filtering corresponds to a top hat
filter in $k$-space, that is, adding $\delta_k$ for larger and larger values
of $k$.
This produces a random walk for each point $\delta=\sum_0^k \delta_k$.  One
can relate $k \rightarrow $R$, $M$ \rightarrow \sigma$ and thus one
talks about trajectories as a function of $M$ or $\sigma$.  Those
trajectories which have crossed some critical value of $\delta=
\delta_c(t)$ at any given time are taken to be collapsed halos.  In
terms of the random walk picture, one can consider $\delta_c$ as an
absorbing barrier and the mass function as given by the distribution
of trajectories in $k$ as they are absorbed.  The resulting number
density is Eq.~(\ref{nps}), 
\eq
N_{PS}(M,t) dM =
\frac{1}{\sqrt{2\pi}} \frac{\rho_0}{M}
\frac{\delta_c(t)}{\sigma^3(M)} \left|\frac{d \sigma^2(M)}{dM}\right| 
\exp \left[-\frac{\delta_c^2(t)}{2 \sigma^2(M)}\right] dM \; .
\en
This is the number of halos, for the amount of mass in halos of this
mass (i.e. the number of trajectories), one multiplies this quantity
by $M/\rho_0$. 

This trajectory approach allows one to consider trajectories which
have crossed the collapse threshold (absorbing barrier) at two
different times, corresponding to being part of a halo of one mass at
one time and another mass at another time, $P_1(M_1, t_1|M_2, t_2)$ and
$P_2(M_2, t_2|M_1, t_1)$, in Eqs. (\ref{p1prob},\ref{p2prob}).

These conditional probabilities can be multiplied together to get
joint probabilities: 
\eq
\begin{array}{l}
P_1(M_1, t_1|M_2, t_2) N_{PS}(M_2,t_2) \times \frac{M_2}{\rho_0} dM_1 dM_2 = \\
P_2(M_2, t_2|M_1, t_1) N_{PS}(M_1,t_1) \times \frac{M_1}{\rho_0}dM_1 dM_2
= P(M_1,t_1,M_2,t_2) dM_1 dM_2
\end{array}
\en
which obey,\footnote{A crucial identity for this is 
\eq
\int_0^\infty du\ e^{-a/u^2 - b u^2} = 
\sqrt{\frac{\pi}{4 b}} e^{-2\sqrt{a b}}
\en
} e.g.,
\eq
dM_1\int dM_2 P_1(M_1, t_1|M_2,t_2) P_1(M_2,t_2|M_3,t_3) =
P_1(M_1,t_1|M_3,t_3) dM_1 \; .
\en
\cite{sethi}.

\subsection{Integration of merger rate}

The starting integral is
\eq 
\begin{array}{ll}
N_{{\rm mer}_{direct}}(M_2, t_1,t_2) =& \int_{t_1}^{t_2} 
dt \int d(M_1 + \Delta M_1) \int dM_1  N(M_1,t) \\
& \times \frac{d P_2}{dt}(M_1 \rightarrow M_1 + \Delta M_1;t) 
P_2(M_2,t_2| M_1 + \Delta M_1, t) \; .
\end{array}
\en
Writing $\sigma^2(M_1 + \Delta M_1) = \sigma_i^2$ and integrating
between times $t_1$ and $t_2$ corresponding to $\delta_c(t_i) \equiv
\delta_i$ one gets 
\eq
\int \frac{d M_1}{M_1}\frac{d \sigma_1^2}{d M_1} \int \frac{d \sigma_i^2}
{2 \pi}
\frac{1}{(\sigma_1^2 - \sigma_i^2)^{3/2}} 
\frac{1}{(\sigma_i^2 - \sigma_2^2)^{1/2}} 
\left[
1-e^{-\frac{\phantom{2}(\delta_1-\delta_2)^2}{2(\sigma_i^2 - \sigma_2^2)}}
\right]
\en
Consider the integral over $\sigma_i^2$:
\eq
\begin{array}{l}
\int \frac{d \sigma_i^2}
{2 \pi}
\frac{1}{(\sigma_1^2 - \sigma_i^2)^{3/2}} 
\frac{1}{(\sigma_i^2 - \sigma_2^2)^{1/2}} 
\left[
1 - e^{-\frac{\phantom{2}(\delta_1-\delta_2)^2}{2(\sigma_i^2 - \sigma_2^2)}}
\right] 
\\ = 
\int_{x_1}^{x_2} dx \frac{1}{x^{1/2} (y-x)^{3/2}} [1- e^{-A^2/(2 x)}]
\end{array}
\en
where we have defined
\eq
A= \delta_1 - \delta_2 \; , \;
x= \sigma_i^2 - \sigma_2^2 \; , \; \; y = \sigma_1^2 - \sigma_2^2 \; .
\en
We then have
\eq
\int_{x_1}^{x_2} dx \frac{1}{x^{1/2} (y-x)^{3/2}} 
= 2 \frac{\sqrt{x_2}}{y\sqrt{y-x_2}} - 
2 \frac{\sqrt{x_1}}{y\sqrt{y-x_1}} 
\en
For the second term we can further define $ z = 1/x$, to get
\eq
\begin{array}{l}
+ \int_{z_1}^{z_2} dz \; e^{-A^2 \frac{z}{2}} (yz - 1)^{-3/2}
= -\frac{1}{y^{3/2}} \left[
\frac{2 e^{-A^2 \frac{z_2}{2}}}{(z_2 - 1/y)^{1/2}} \right. \\
\phantom{xxxxx}  - 
e^{-\frac{A^2}{2 y}}\sqrt{2 \pi}A 
(1 - {\rm erf}[\frac{A}{2^{1/2}} (z_2 - 1/y)^{1/2}]) 
 \left. - (z_2 \to z_1) \right]
\end{array}
\en

\end{document}